\documentclass[11pt]{amsart}
\usepackage{amssymb}
\usepackage{amsmath,amsthm,amsfonts}  
\usepackage[applemac]{inputenc} 
 \usepackage{graphicx}
\usepackage{amsthm}
\usepackage{pdfsync}
\usepackage{fancyhdr}
\usepackage[T1]{fontenc}

\usepackage{color}
\usepackage{xcolor}
\usepackage[breaklinks,colorlinks,backref]{hyperref}
\hypersetup{
    colorlinks, 
    linktoc=all, 
    linkcolor=red,  
  citecolor=hyptxt,
  urlcolor=blue}
  \hypersetup{
  citebordercolor=,
  filebordercolor=green,
  linkbordercolor=blue
}
\definecolor{hyptxt}{rgb}{0.7, 0.4, 0.9}
\usepackage{bibtopic}

\newtheorem{prop}{Proposition}[section]

\newtheorem{coro}{Corollary}[section]

\newcommand{\beprop}{\begin{prop}}
\newcommand{\enprop}{\end{prop}}
\newcommand{\bprf}{\begin{proof}}
\newcommand{\eprf}{\end{proof}}

\definecolor{hervecolor}{rgb}{0.8,0,0.7}
     
\newcommand{\ket}[1]{|\kern.3ex#1\kern.3ex\rangle}
\newcommand{\bra}[1]{\langle\kern.3ex #1 \kern.3ex|}
\newcommand{\scalar}[2]{\langle\kern.3ex #1 \kern.3ex|\kern.3ex#2\kern.3ex\rangle}

\newcommand{\ii}{\mathsf{i}}

\def\calH{{\mathcal H }}

\def\R{\mathbb{R}}
\def\N{\mathbb{N}}

\def\lg{\langle }
\def\rg{\rangle }

\def\vap{\varpi}

\def\ud{\mathrm{d}}

\def\sfM{\mathsf{M}}

\def\sfMv{\mathsf{M}^{\vap}}
\def\cMv{\mathcal{M}^{\vap}}

\def\sfMa{\sfM^{a\mathcal{W}}}
\def\vapa{\vap_{a\mathcal{W}}}

\def\Om{\Omega}

\def\eal{e^{(\alpha)}}
\def\Lal{L^{(\alpha)}}
\def\cdm{\mathsf{C}_{\mathrm{DM}}}

\numberwithin{equation}{section}

\title{Covariant Affine Integral Quantization(s)}
\author{Jean Pierre Gazeau and Romain Murenzi}
\address{Laboratoire APC, Universit\'e Paris 7-Denis Diderot, 10, rue A. Domon et L. Duquet, 75205 Paris Cedex13, France}\email{gazeau@apc.univ-paris7.fr}
\address{The World Academy of Sciences, TWAS/ICTP\\
Via Costiera 1, Trieste 34151, Italy
}
\email{rmurenzi@twas.org}

\date{\today}        

\begin{document}

\begin{abstract}
Covariant affine integral quantization of the half-plane  is studied and applied to the motion of a particle on the half-line. We examine the consequences of different  quantizer operators built from weight functions on the half-plane. To illustrate the procedure, we examine two particular choices of the weight function, yielding  thermal density operators and affine inversion respectively. The former gives rise to a temperature-dependent probability distribution on the half-plane whereas the  later yields the usual canonical quantization  and a quasi-probability distribution (affine Wigner function) which is real, marginal in both momentum $p$ and position $q$. 
\end{abstract}
\maketitle

\tableofcontents

\section {Introduction}
\label{intro}

This work is a continuation of previous studies devoted to affine integral quantization of the half-plane \cite{bergaz14,aagbook13,balfrega14,gazhell15} and its applications to early or quantum cosmology \cite{berdagama14,berdagamal15,beczgamapie15,beczgamapie15A,beczgama15A,beczgama15B}. In the latter works, the method was based on the use of affine coherent states or wavelets. Here we extend it to a class of operators including density operators and the inversion operator allowing a sort of Wigner transform for the affine group.

The coherent state quantization which was used in the applications cited above is a particular approach pertaining to
what is named in \cite{bergaz14} integral quantization. When a
group action is involved in the construction, one can insist on
covariance aspects of the method. A detailed presentation of the
procedure is given in \cite{bergaz14} and in Chapt.\;11 of
\cite{aagbook13}. In the case of  the Weyl-Heisenberg group, weight functions defined on the euclidean plane viewed as a phase space were at the heart of the construction of the covariant integral quantization.  In a certain sense these  functions, or their symplectic Fourier transforms,   correspond to the Cohen ``$f$'' function \cite{cohen66} (for more details see \cite{cohen13} and references therein) or to Agarwal-Wolf filter functions \cite{agawo70}, even though  these authors were not directly proposing new quantization procedures.

The organisation of the paper is as follows. In Section \ref{covintquant}, we give a short compendium of covariant integral quantization  approach. All necessary details on the affine group of the real line are given in Section \ref{affG}, geometry, unitary irreducible representation (UIR),  UIR matrix elements in a particular Laguerre Hilbertian basis from which is built a specific density operator called thermal state. 
In Section \ref{weight} we show how to build symmetric  operators as UIR transform of  weight functions on the half-plane. 
 The affine transports of these operators allow  a resolution of the identity, a necessary ingredient for quantization. The main results concerning these affine  covariant quantizations are presented in Section \ref{genform}. In 
Section \ref{QaffCSth} we particularise in Subsection  \ref{QaffCS} this quantization method  by using one-rank density operators, i.e. affine coherent states (ACS), recalling results given in previous articles, and we also consider in 
Subsection \ref{thermal} the quantization issued from Laguerre thermal states.  
Section \ref{inversion} is devoted to the quantization  issued from the affine inversion operator which correspond to a specific weight function and which leads to a Wigner-like quasi-distribution on the half-plane. To a certain extent, the results described in this  section are related to previous works, particularly the pioneering contributions by J. Bertrand and P. Bertrand \cite{berber92,berber98}, the more mathematically oriented articles by Ali \textit{et al} \cite{alatchuwo00,alfukra03} and  by the comprehensive  approach by Gayral \textit{et al} \cite{gaygrava08} with references therein, devoted to Fourier analysis on the (full) affine group, its Stratonovich–Weyl quantizer, and the corresponding Wigner functions. The application to the half-oscillator is given as an elementary example and graphical illustrations are given to compare the Wigner quasi-distribution with probability distributions issued from ACS. For another interesting approach, based on representations of  the larger group  SL$(2, \R)$ and its covering(s),  to the quantization of the motion of a particle moving on a half-line with ‘hard wall’ boundary condition, see the recent PhD thesis of Jung \cite{jung12}. 
We conclude in   Section \ref{conclu} with comments and a brief description of our future work on the same subject.  

\section{Covariant integral quantizations} \label{covintquant}

Lie group representations \cite{barracz77} offers a wide range of
possibilities for implementing integral quantization(s).  Let $G$
be a Lie group with left Haar measure $\mathrm{d}\mu(g)$, and let
$g\mapsto U\left(g\right)$ be a unitary irreducible representation
(UIR) of $G$ in a Hilbert space $\mathcal{H}$. Let us consider a bounded
operator $\mathsf{M}$ on $\mathcal{H}$ and suppose that the
operator
\begin{equation}
\label{boundR}
\mathsf{R}:=\int_{G}\mathsf{M}\left(g\right)\mathrm{d}\mu\left(g\right),\
\mathsf{M}\left(g\right):=U\left(g\right)\mathsf{M}\,U\left(g\right)^{\dagger}\,,
\end{equation}
is defined in a weak sense. From the left invariance of
$\mathrm{d}\mu(g)$ we have
\begin{equation}
\label{comRU}
U\left(g_{0}\right)\mathsf{R}\,U^{\dagger}\left(g_{0}\right)=\int_{G}\mathsf{M}\left(g_{0}g\right)
\mathrm{d}\mu\left(g\right)=\mathsf{R}\,,
\end{equation}
so $\mathsf{M}$ commutes with all operators $U(g)$, $g\in G$. Thus, from
Schur's Lemma, $\mathsf{M}=c_{\mathsf{M}}I$ with
\begin{equation}
\label{cM}
c_{\mathsf{M}}=\int_{G}tr\left(\rho_{0}\mathsf{M}\left(g\right)\right)\mathrm{d}\mu\left(g\right)\,,
\end{equation}
where the unit trace non-negative operator ($\sim$ density operator) $\rho_{0}$ is chosen in
order to make the integral converge. This family of operators
provides the resolution of the identity on $\mathcal{H}$.
\begin{equation}
\int_{G}\mathsf{M}\left(g\right)\mathrm{d}\nu\left(g\right)=I,\qquad
\mathrm{d}\nu\left(g\right)
:=\frac{\mathrm{d}\mu\left(g\right)}{c_{\mathsf{M}}}\,.\label{eq:resolution}
\end{equation}
and the subsequent quantization of complex-valued functions (or
distributions, if well-defined) on $G$
\begin{equation}
\label{quantgr} f\mapsto A_{f}=\int_{G}\, \mathsf{M}(g)\,
f(g)\,\mathrm{d}\nu(g)\,,
\end{equation}
This linear map, function $\mapsto$ operator in $\mathcal{H}$, is
covariant in the sense that
\begin{equation}
\label{covarG} U(g)A_{f}U(g)^{\dagger}=A_{\mathfrak{U}_{r}(g)f}\,.
\end{equation}
In the case when $f\in L^{2}(G,\mathrm{d}\mu(g))$, the action
$(\mathfrak{U}_{r}(g)f)(g^{\prime}) :=f(g^{-1}g^{\prime})$ defines the regular
representation of $G$.

The other face of integral quantization concerns a consistent semi-classical analysis of the operator $A_f$ based on it. It  is implemented
through the study of the so-called lower symbols. Suppose that $\mathsf{M}$ is
a density operator $\mathsf{M}=
\rho$ on $\mathcal{H}$. Then the operators $\rho(g)$ are also
density, and this allows to build a new function $\check{f}(g)$, called lower symbol,  as
 \begin{equation}
\label{lowsymb} \check{f}(g)\equiv \check{A_f}:=\int_{G}\,
tr(\rho(g)\,\rho(g^{\prime}))\,
f(g^{\prime})\mathrm{d}\nu(g^{\prime})\,.
\end{equation}
The map $f\mapsto \check{f}$ is a generalization of the Berezin or
heat kernel transform on $G$ (see \cite{hall06} and references
therein). Choosing for $\mathsf{M}$ a density operator $\rho$ has multiple advantages, peculiarly
in  regard to probabilistic aspects both on classical and quantum levels \cite{gazhell15}. 

Let us illustrate the above procedure with the case of square
integrable UIR's and rank one $\rho$. For a square-integrable UIR
$U$ for which $\left\vert \psi\right\rangle $ is an admissible
unit vector, i.e.,
\begin{equation}
\label{cpsi} c(\psi):=\int_{G}\mathrm{d}\mu(g)\,|\left\langle
\psi\right\vert U\left(g\right)\left\vert \psi\right\rangle
|^{2}<\infty\,,
\end{equation}
the resolution of the identity is obeyed by the coherent states
$\left\vert \psi_{g}\right\rangle =U(g)\left\vert
\psi\right\rangle$, in a generalized sense, for the group $G$:
\begin{equation}
\label{residsq} \int_{G}\rho(g)  \mathrm{d}\nu(g)=I\
,\quad \mathrm{d}\nu(g)=
\frac{\mathrm{d}\mu(g)}{c(\psi)}\, , \quad
\rho(g)=\left\vert \psi_g\right\rangle \left\langle
\psi_g\right\vert \,.
\end{equation}

\section{The affine group and its representation $U$} 
\label{affG}

\subsection{Half-plane and affine group}

The half-plane can be viewed as the phase
space for the (time) evolution a positive physical quantity such as the position of a particle moving in the half-line, or, at the opposite, the contracting-expanding volume of the Universe. Let  the
upper half-plane $\Pi_{+}:=\{(q,p)\,|\, p\in\mathbb{R}\,,\, q>0\}$
be equipped with the  measure $\mathrm{d}q\mathrm{d}p$. Together
with the multiplication
\begin{equation}
\label{multaff} (q,p)(q_{0},p_{0})=(qq_{0},p_{0}/q+p),\,
q\in\mathbb{R}_{+}^{\ast},\, p\in\mathbb{R}\, ,
\end{equation}
the unity $(1,0)$ and the inverse
\begin{equation}
\label{invaff} (q,p)^{-1}= \left(\frac{1}{q}, -qp \right)\, ,
\end{equation}
$\Pi_{+}$ is viewed as the affine group Aff$_{+}(\mathbb{R})$ of
the real line, and the  measure $\mathrm{d}q\mathrm{d}p$ is
left-invariant with respect to this action.

\subsection{Representation}
The affine group
Aff$_{+}(\mathbb{R})$ has two non-equivalent UIR $U_{\pm}$
\cite{gelnai47,aslaklauder68}. Both are square integrable and this
is the rationale behind \textit{continuous wavelet analysis} \cite{wave1,wave2,wave3,aagbook13}. Only the UIR $U_{+}\equiv U$ is concerned in the rest of the paper. This representation is realized
in the Hilbert space
$\mathcal{H}=L^{2}(\mathbb{R}_{+}^{\ast},\mathrm{d}x)$ as 
\begin{equation}
U(q,p)\psi(x)=(e^{\ii px}/\sqrt{q})\psi(x/q)\,.\label{affrep+}
\end{equation}
The most immediate (and well-known) orthonormal basis of $L^{2}(\mathbb{R}_{+}^{\ast},\mathrm{d}x)$ is that one which is  built from Laguerre polynomials,
\begin{equation}
\label{LagOB}
 e^{(\alpha)}_n(x) := \sqrt{\frac{n!}{(n+\alpha)!}}\, e^{-\frac{x}{2}}\, x^{\frac{\alpha}{2}}\, L_n^{(\alpha)}(x)\, , \ \int_{0}^{\infty}\eal_n(x)\, \eal_{n^{\prime}}(x) \ud x = \delta_{n n^{\prime}}\,, 
\end{equation}
where $\alpha > -1$ is a free parameter, and $(n+\alpha)! = \Gamma(n+\alpha + 1)$. Actually,  since we wish to work with  functions which, with a certain number of their derivatives,  vanish at the origin, the parameter $\alpha$ should  be imposed to be larger than  some  $\alpha_0 >0$. 

The matrix elements  $U^{(\alpha)}_{m n}  (q,p):=\lg \eal_m|U(q,p)|\eal_n\rg$ of the representation $U$ with respect to this basis are given in terms of Gauss hypergeometric polynomial or Jacobi polynomial \cite{magnus66} by 
\begin{align}
\label{intform} U^{(\alpha)}_{m n}  (q,p)  &= \frac{1}{q^{(\alpha+1)/2}}\, \sqrt{\frac{m!n!}{(m+\alpha)!(n+\alpha)!}} \int_{0}^{\infty} \ud x\, e^{-\left(\frac{1}{2} + \frac{1}{2q}-\ii p \right)x}\, x^{\alpha}\, \Lal_m(x)\, \Lal_n\left(\frac{x}{q}\right)\\
\nonumber  & =  2^{\alpha+1} \,\sqrt{\binom{m+n+ \alpha}{m}\binom{m+n+ \alpha}{n}} q^{(\alpha+1)/2}\frac{ Z_-^{m}\bar Z_-^{n}}{\bar Z_+^{m+n+\alpha+1}}\times\\ 
\nonumber     &\times  {}_2F_1 \left(-m,-n;-m-n-\alpha; \left\vert\frac{Z_+}{Z_-}\right\vert^2\right)\\
  \label{matelU2q} &=  2^{\alpha+1}\, \sqrt{\frac{(n+\alpha)!}{(m+\alpha)!}\,  \frac{m!}{n!}}\, q^{(\alpha+1)/2}\, \frac{\bar Z_-^{n-m}Z^m_+}{\bar Z_+^{n+\alpha +1}} 
   P_m^{(n-m,\alpha)}(Y)\, ,
\end{align}
where 
\begin{equation}
\label{matelU3q} Z_{\pm}:= q\pm 1 + 2\ii qp\,,    \quad Y:= 1 - 2\left\vert\frac{Z_-}{Z_+}\right\vert^2\, . 
\end{equation}
The integral formula involving associated Laguerre polynomials is found in \cite{gradryz07}, 850-4.
 One easily verifies the unitarity
\begin{equation}
\label{unitarity}
U^{(\alpha)}_{mn}(q^{-1},-qp)= {U^{(\alpha)}}_{mn}^{-1}(q,p)= (U^{(\alpha)})^{\dag}_{mn}(q,p)=  \overline{U^{(\alpha)}_{nm}(q,p)} 
\end{equation}
from $Y(q,p) = Y(1/q,-qp)$ and \cite{magnus66}
\begin{equation*}
 P_m^{(n-m,\alpha)}(X) = \frac{(m+\alpha)!}{(n+\alpha)!}\,  \frac{n!}{m!}\,\left(\frac{X-1}{2}\right)^{m-n}\, P_n^{(m-n,\alpha)}(X)\, .  
\end{equation*}
These matrix elements obey orthogonality relations for the affine group. Since this group is not unimodular, there exists a positive self-adjoint and invertible operator 
$\mathsf{C}_{\mathrm{DM}}$, the Duflo-Moore operator, such that \cite{aagbook13}
\begin{equation}
\label{orthaffine}
\int_{\Pi_{+}} \mathrm{d}q\,\mathrm{d}p \lg U(q,p)\psi| \phi\rg \, \overline{\lg U(q,p)\psi^{\prime}| \phi^{\prime}\rg}= \lg \mathsf{C}_{\mathrm{DM}}\psi| \mathsf{C}_{\mathrm{DM}}\psi^{\prime}\rg\, \lg\phi^{\prime}|\phi\rg\, , 
\end{equation}
for any pair $(\psi,\psi^{\prime})$ of admissible vectors, i.e. which obey $\Vert \mathsf{C}_{\mathrm{DM}}\psi\Vert < \infty$,
$\Vert \mathsf{C}_{\mathrm{DM}}\psi^{\prime}\Vert < \infty$, 
and any pair $(\phi,\phi^{\prime})$ of vectors in $L^{2}(\mathbb{R}_{+}^{\ast},\mathrm{d}x)$. For the affine group, the Duflo-Moore operator is the multiplication operator 
\begin{equation}
\label{CDM}
\mathsf{C}_{\mathrm{DM}}\psi(x) := \sqrt{\dfrac{2\pi}{x}}\psi(x)\equiv \sqrt{\dfrac{2\pi}{Q}}\psi(x)\, , 
\end{equation}
where $Q \psi(x) := x \psi(x)$ is the basic positive self-adjoint multiplication operator with $|x\rg= \delta_x$ as eigendistributions.  Thus, the admissibility condition for $\psi \in L^{2}(\mathbb{R}_{+}^{\ast},\mathrm{d}x)$ amounts to \begin{equation}
\label{admcond}
\Vert \mathsf{C}_{\mathrm{DM}}\psi\Vert^2=2 \pi \int_0^{+\infty}\frac{\ud x}{x} \vert \psi(x)\vert^2 < \infty\, . 
\end{equation}
For the sequel, due to the non-unimodular nature of the affine group, that operator $\cdm$ is  expected to play an important r\^ole. A first important and direct consequence of Eq\;\eqref{orthaffine} is the resolution  of the identity in $L^{2}(\mathbb{R}_{+}^{\ast},\mathrm{d}x)$:
\begin{equation}
\label{residCDM}
\frac{1}{\Vert \mathsf{C}_{\mathrm{DM}}\psi\Vert^2}\int_{\Pi_{+}} \mathrm{d}q\,\mathrm{d}p \, U(q,p)|\psi\rg\lg \psi|U^{\dag} (q,p)= I\, . \end{equation}
satisfied by the family  of \textit{affine coherent states} (ACS) $U(q,p)|\psi\rg \equiv |q,p\rg$  built from  the  admissible vector $\psi$ through the unitary affine transport $U$.

 Note that the equation \eqref{orthaffine} implies for the matrix elements the integral formula,
\begin{equation}
\label{intmatel}
\int_{\Pi_{+}} \mathrm{d}q\,\mathrm{d}p\, \overline{U^{(\alpha)}_{mn}(q,p)}\, U^{(\alpha)}_{m^{\prime}n^{\prime}}(q,p)= 2\pi\, \delta_{mm^{\prime}}\,
\lg  \eal_{n}|(1/Q)\eal_{n^{\prime}}\rg\, .
\end{equation}

Let us end this section with a formula giving the trace of the operator $U(q,p)$ from \cite{magnus66}.
\begin{equation}
\label{traceU}
\mathrm{Tr}\;U(q,p) = \sum_{m=0}^{\infty}U_{mm}^{(\alpha)}(q,p)= \frac{\sqrt{q}}{\vert q - 1\vert}\, .
\end{equation}

\subsection{The  ``thermal state'' case}

Here we consider the temperature-dependent density operator 
 \begin{equation}
\label{plboltrhohp}
\rho_t= \left( 1- t\right)\sum_{n=0}^{\infty} t^n|e_n \rg\lg e_n|\,, \quad t = e^{-\tfrac{\hbar \omega }{k_B T}}\, ,  
\end{equation}  
where $\{| e_n\rg\, |\, n\in \mathbb{N}\}$ is an orthonormal basis of $\mathcal{H}$. By choosing the Laguerre orthonormal basis \eqref{LagOB} for which $\rho_t\equiv \rho_t^{(\alpha)}$, we find  from \cite{magnus66} that the operator $\rho_t$ acts on $\mathcal{H}=L^{2}(\mathbb{R}_{+}^{\ast},\mathrm{d}x)$ as the integral transform
\begin{equation}
\label{intransf}
\rho^{(\alpha)}_t: \psi(x) \mapsto \rho^{(\alpha)}_t(\psi)(x) = \int_{0}^{\infty} \mathcal{K}^{(\alpha)}_t(x,y)\,\psi(y)\, \ud y\, , 
\end{equation}
where the integral kernel is given by
\begin{equation}
\label{intkerLag}
\mathcal{K}^{(\alpha)}_t(x,y) = t^{-\alpha/2}\, e^{-\frac{1}{2}\frac{1+t}{1-t}(x+y)}\, I_{\alpha}\left(2\frac{\sqrt{t x y}}{1-t}\right)\, . 
\end{equation}
This closed form is obtained from the expression of the Poisson kernel for the Laguerre polynomials in terms of a modified Bessel function, 8.976--1 in  \cite{gradryz07} 
\begin{equation}
\label{lagpoisson}
\sum_{n=0}^{\infty}\frac{n!}{(n+\alpha)!} \,\Lal_n(x)\,\Lal_n(y) \, t^n= \frac{1}{(xyt)^{\alpha/2}(1-t)}\,e^{-(x+y)\frac{t}{1-t}}\, I_{\alpha}\left(\frac{2\sqrt{xyt}}{1-t}\right)\, , 
\end{equation}
for $\vert t\vert <1$. 
Again, one  derives from the Schur lemma  that
\begin{equation}
\label{rhoTqp}
\rho^{(\alpha)}_t(q,p):= U(q,p)\rho^{(\alpha)}_tU(q,p)^\dagger
\end{equation}
 resolves the identity, 
\begin{equation}
\label{residrhoTqp}
\int_{\Pi_{+}}\rho^{(\alpha)}_t(q,p) \,\frac{\mathrm{d}q\,\mathrm{d}p}{c_{\rho^{(\alpha)}_t}}= I\, , 
\end{equation}
where the constant $c_{\rho^{(\alpha)}_t}$ is obtained  through standard calculations in wavelet theory,
\begin{align}
\label{croexpl}
\nonumber c_{\rho^{(\alpha)}_t} &= \int_{\Pi_{+}}\lg \eal_0| \rho_t(q,p)|\eal_0\rg \,\mathrm{d}q\,\mathrm{d}p\\
\nonumber & = (1-t)\sum_{n=0}^{\infty} t^n\,\int_{\Pi_{+}}\lg \eal_0|U(q,p|\eal_n\rg
\lg\eal_n |U(1/q,-qp)\rg\,\mathrm{d}q\,\mathrm{d}p\\
\nonumber &= 2\pi(1-t)\sum_{n=0}^{\infty}\frac{n!}{\alpha!(n+\alpha)!}\, t^n\, \int_0^{\infty} \ud q \int_0^{\infty} \ud x e^{-(1+q)x} \, x^{2\alpha}\, q^{\alpha}\left(\Lal_n(x)\right)^2\\
\nonumber &= 2\pi(1-t) \int_0^{\infty} \frac{\ud x}{x} e^{-x} \, x^{\alpha}\, \sum_{n=0}^{\infty}\frac{n!}{(n+\alpha)!}\, t^n\, \left(\Lal_n(x)\right)^2\\
\nonumber &= 2\pi \,t^{-\alpha/2} \,\int_0^{\infty} \frac{\ud x}{x} \, e^{-\frac{1+t}{1-t}x}\, I_{\alpha}\left(\frac{2\sqrt{t}}{1-t}x\right)\\
&= \frac{2\pi}{\alpha}\, . 
\end{align}
Here we have used the integral formula involving a Bessel function:
\begin{equation}
\label{mbessint}
\int_{0}^{\infty}\frac{\ud x}{x}\, e^{-\gamma x}\, I_{\alpha}(\mu x) = \frac{1}{\alpha}\left[ \frac{\gamma}{\mu} - \sqrt{\frac{\gamma^2}{\mu^2}-1}\right]^{\alpha}\, . 
\end{equation} 
Thus, the resolution of the identity imposes the painless restriction $\alpha >0$ and reads finally
\begin{equation}
\label{residrhoTqpF}
\alpha\int_{\Pi_{+}}\rho^{(\alpha)}_t(q,p) \,\frac{\mathrm{d}q\,\mathrm{d}p}{2\pi}= I\, .
\end{equation}

\section{Symmetric operators from weight functions}
\label{weight}
A general method to get density or more general (symmetric) bounded operators is the following.  
Let us  choose like in \cite{balfrega14} (see also \cite{bergaz14} for the Weyl-Heisenberg group) a suitably
localized weight function $\mathsf{\varpi}(q,p)$ on the half-plane
such that the integral 
\begin{equation}
\int_{\Pi_{+}}\cdm^{-1} U(q,p)\cdm^{-1} \,\varpi(q,p)\,\mathrm{d}q\,\mathrm{d}p:=\sfMv\label{affbop}
\end{equation}
defines a symmetric operator.  The non-unimodularity of the affine group justifies the double presence of the inverse of the Duflo-Moore operator, at the difference of the situation we had for the Weyl-Heisenberg group in \cite{bergaz14}.  From the symmetric property  $\sfMv={\sfMv}^{\dagger}$,
we find that the weight function must satisfy 
\begin{equation}
\varpi(q,p)=\frac{1}{q}\overline{\varpi\left(\frac{1}{q},-qp\right)}\,.\label{eq:condition1}
\end{equation}
Trivial (but with not so trivial consequences!) solutions are 
\begin{align}
\label{trivsol1}
\varpi(q,p)    &= \frac{1}{\sqrt{q}}\, ,    \\
\label{trivsol2}  \varpi(q,p)    &=  e^{\pm \ii \sqrt{q}p}\,. 
\end{align}
These two elementary solutions combine to yield the self-adjoint nilpotent inversion operator $\mathcal{I}$ defined on $L^2(\R_+^*,\ud x)$:
\begin{equation}
\label{invop}
\int_{\Pi_{+}}\cdm^{-1} U(q,p)\cdm^{-1} \,\frac{e^{-\ii\sqrt{q}p}}{2\sqrt{q}}\,\mathrm{d}q\,\mathrm{d}p = \mathcal{I}\, , \quad (\mathcal{I}\psi)(x): = \frac{1}{x}\, \psi\left(\frac{1}{x}\right)\, , \quad \mathcal{I}^2 = I\, . 
\end{equation}
Note here that the two inverse Duflo-Moore operators simplify to the factor $1/(2\pi)$ since $1/\sqrt{Q}\, \mathcal{I} \,1/\sqrt{Q} = \mathcal{I}$.  Section \ref{inversion} is devoted to the study of this important particular case after including a factor 2 in order to get the unit trace operator
\begin{equation}
\label{vapweyl}
\sfM^{a\mathcal{W}}\equiv 2\mathcal{I} =  \int_{\Pi_{+}}U(q,p)\,\vap_{a\mathcal{W}}(q,p)\,\mathrm{d}q\,\mathrm{d}p\, ,\quad  
\vap_{a\mathcal{W}}(q,p) :=\frac{e^{-\ii\sqrt{q}p}}{\sqrt{q}}\, .
\end{equation}
This operator is the affine counterpart of the operator yielding the Weyl-Wigner integral  quantization when the phase space is $\R^2$, i.e. we  deal with Weyl-Heisenberg  symmetry.  Its unit trace property is  proved below. 

Let us now establish a necessary  condition on $\vap$ to have a unit trace operator $\sfMv$.
\beprop
\label{neccondtr1}
Suppose that the operator $\sfMv$ is unit trace class $\mathrm{Tr}(\sfMv) = 1$. Then  its corresponding  weight function $\vap(q,p)$  obeys
\begin{equation}
\label{Tr1cond}
  \frac{\vap(1,0)}{2} +\frac{ \ii}{2\pi}\, \underset{\epsilon \to 0}{\lim}\int_{\R/[-\epsilon,\epsilon]}\frac{\vap(1,p)}{p}\,\ud p = 1\, . 
\end{equation}
\enprop
\bprf
From the definition \eqref{affbop} of the operator $\sfMv$ the first step of the proof consists in determining the trace of the operator $\cdm^{-1} U(q,p)\cdm^{-1}$. For that we use the resolution of the identity \eqref{residCDM} provided by an admissible vector $\psi$. We choose here a real $\psi$ for convenience. We have successively
\begin{align}
\label{}
 \nonumber \mathrm{Tr}\left(\cdm^{-1} U(q,p)\cdm^{-1}\right)  =& \frac{1}{\Vert \mathsf{C}_{\mathrm{DM}}\psi\Vert^2}\int_{\Pi_{+}} \mathrm{d}q^{\prime}\,\mathrm{d}p^{\prime}\times\\
 \nonumber \times & \lg \psi|U^{\dag}(q^{\prime},p^{\prime})\, \cdm^{-1} U(q,p)\cdm^{-1}\,U (q^{\prime},p^{\prime})\,|\psi\rg   \\
\nonumber   = &  \frac{1}{2\pi q\Vert \mathsf{C}_{\mathrm{DM}}\psi\Vert^2}\int_0^{+\infty}\frac{ \mathrm{d}q^{\prime}}{q^{\prime}}\int_0^{+\infty}\ud x \, x\, e^{\ii px}\, \psi\left(\frac{x}{q^{\prime}}\right) \, \psi\left(\frac{x}{qq^{\prime}}\right)\times\\
\nonumber \times & \int_{-\infty}^{+\infty}\ud p^{\prime}\, e^{-\ii p^{\prime}x\left(\frac{q-1}{q}\right)}\,. 
\end{align}
Integrating on $p^{\prime}$ gives $2\pi \delta\left(x\frac{q-1}{q}\right)= 2\pi \frac{q^2}{x}\, \delta(q-1)$, which allows to fix $q=1$. Then the change of variable $q^{\prime} \mapsto y= x/q^{\prime}$ allows to separate the two remaining integrals. That one on $(\psi(y))^2/y$ yields  $\Vert \mathsf{C}_{\mathrm{DM}}\psi\Vert^2/(2\pi)$ and the last one is the inverse Fourier transform of the (Heaviside) step function which is equal to $\frac{1}{\sqrt{2\pi}}\left(\pi\delta(p) +\ii\, \mathrm{p}.\mathrm{v}.\frac{1}{p}\right)$, where $\mathrm{p}. \mathrm{v}.$ denotes the Cauchy principal value. Finally, 
\begin{equation}
\label{tracecdmU}
 \mathrm{Tr}\left(\cdm^{-1} U(q,p)\cdm^{-1}\right)= \frac{1}{2}\,\delta(q-1)\left(\delta(p) + \frac{\ii}{\pi}\,\mathrm{p}.\mathrm{v}.\frac{1}{p}\right)\, .
\end{equation}
\eprf
Note that the condition \eqref{Tr1cond} can be also written as
\begin{equation}
\label{Tr2cond}
\mathrm{Tr}(\sfMv)= \frac{1}{\sqrt{2\pi}}\int_0^{+\infty}\ud x \,\hat\vap_p(1,-x)= 1\, ,
\end{equation}
where $\hat{\vap}_p$ is the partial Fourier transform of $\vap$ with respect to the variable $p$ defined as
\begin{equation}
\label{parcoure} \hat{\vap}_p(q,x)=
\frac{1}{\sqrt{2\pi}}\int_{-\infty}^{+\infty} e^{-\ii px} \vap(q,p)\, .
\end{equation}
From the Dirichlet integral $\int_0^{+\infty}\mathrm{sinc}t\,\ud t = \frac{\pi}{2}$, we easily check that the special weight $\vap_{a\mathcal{W}}(q,p)$ given in Eq.\;\eqref{vapweyl}  satisfies the condition
\eqref{Tr1cond}.

Suitably weighted and scaled diagonal matrix elements $U^{(\alpha)}_{mm}(q,p)$ are example of solutions to the functional equation \eqref{eq:condition1}. Indeed, let us define 
\begin{equation}
\label{diagosol}
\varpi^{(\alpha)}_m(q,p):= \frac{1}{\sqrt{q}}\, U^{(\alpha)}_{mm}(q,p)= \frac{2^{\alpha+1}\, q^{\alpha/2}}{(q+1 +2\ii qp)^{\alpha +1}}\, e^{2\ii m \theta(q,p)}
   P_m^{(0,\alpha)}(Y(q,p)) 
\, ,
\end{equation}
where $\theta(q,p) = \arg(q+1 + 2\ii qp)$. 
From \eqref{unitarity}, one easily checks that the $s$-dependent weight function defined by
\begin{equation}
\label{pioms}
\varpi^{(\alpha)}_{m;s}(q,p):= \varpi^{(\alpha)}_m(q,sp)\, ,  
\end{equation}
verifies \eqref{eq:condition1} for any $s\in \R$ and $m\in \N$.  
With this function in  hand we have the following interesting result.
\beprop
Let $\mathsf{P}_m= |\eal_m\rg\lg\eal_m|$ be the orthogonal projector on the Laguerre basis element $\eal_m(x)$ and let $\rho^{(\alpha)}_{m;s}$ be the self-adjoint operator defined by
\begin{equation}
\label{rhoms}
\rho^{(\alpha)}_{m;s} = \int_{\Pi_{+}}U(q,p)\,\varpi^{(\alpha)}_{m;s}(q,p)\,\mathrm{d}q\,\mathrm{d}p\, , \ \mbox{with}\ s>0\, . 
\end{equation}
Then we have
\begin{equation}
\label{projrhoms}
\sqrt{Q}\, U^{\dag}(s,0)\rho^{(\alpha)}_{m;s}\,U(s,0)\,  \sqrt{Q} = 2\pi \, \mathsf{P}_m\, , 
\end{equation}
or equivalently
\begin{equation}
\label{projrhoms2}
 \mathsf{P}^{(\alpha)}_m= \int_{\Pi_{+}}\cdm^{-1}\,U(q,sp)\,\cdm^{-1}\,\varpi^{(\alpha)}_{m;s}(q,p)\,\mathrm{d}q\,\mathrm{d}p\, .
\end{equation}
\enprop
\bprf
Let us first apply \eqref{rhoms} to $\psi \in L^2(\R^*_+,\ud x)$ while keeping the integral form \eqref{intform} of $U^{(\alpha)}_{mm}$. 
\begin{align}
\label{}
\nonumber  \left(\rho^{(\alpha)}_{m;s}\psi\right)(x)  & =\int_{\Pi_{+}}\mathrm{d}q\,\mathrm{d}p \, U(q,p)\, \psi(x)\,\frac{1}{\sqrt{q}}\,U^{(\alpha)}_{mm}(q,sp) \\
\nonumber    & = \frac{m!}{(m+\alpha)!} \int_{\Pi_{+}}\frac{\mathrm{d}q \,\mathrm{d}p}{q^{\frac{\alpha +3}{2}}} \int_{0}^{\infty}\ud y \, y^{\alpha}\, 
    e^{-\left(\frac{1}{2}+ \frac{1}{2q} -\ii sp\right)y +ipx}\times\\
    &\times \Lal_m(y)\,\Lal_m\left(\frac{y}{q}\right)\, \psi\left(\frac{x}{q}\right)\, .
\end{align}
After performing the integration on $p$, which gives the Dirac $2\pi\delta(x-sy)$, and then integrating on $y$, one obtains
\begin{equation}
\label{interint}
\left(\rho^{(\alpha)}_{m;s}\psi\right)(x)= \frac{2\pi\,m!}{(m+\alpha)!}\,e^{-\frac{x}{2s}}\, \left(\frac{x}{s}\right)^{\alpha}\, \Lal_m\left(\frac{x}{s}\right) \int_{0}^{\infty}\frac{\ud q}{q^{\frac{\alpha+3}{2}}} \, 
    e^{-\frac{x}{2sq}}\, \Lal_m\left(\frac{x}{sq}\right)\, \psi\left(\frac{y}{q}\right)\, . 
\end{equation}
After performing the change of variable $q\to x/q= u$, we obtain 
\begin{align}
\label{sepvar}
\nonumber \left(\rho^{(\alpha)}_{m;s}\psi\right)(x)&= \frac{2\pi\,m!}{(m+\alpha)!}\,e^{-\frac{x}{2s}}\, \left(\frac{x}{s}\right)^{\frac{\alpha-1}{2}}\, \Lal_m\left(\frac{x}{s}\right)\times\\
\nonumber &\times \int_{0}^{\infty}\frac{\ud u}{s} \, 
    e^{-\frac{u}{2s}}\,  \left(\frac{u}{s}\right)^{\frac{\alpha-1}{2}}\,\Lal_m\left(\frac{u}{s}\right)\, \psi\left(u\right)\\
  &= 2\pi \, \left(U(s,0)\sqrt{\frac{1}{Q}}\, \eal_m\right)(x)\,\left \lg U(s,0)\sqrt{\frac{1}{Q}}\, \eal_m \right|\left.\psi\right\rg\, . 
\end{align}
\eprf
This result is interesting since it allows to give the thermal state $\rho_t$ in \eqref{plboltrhohp} the integral representation
\begin{align}
\label{thermint}
\nonumber \rho^{(\alpha)}_t &= \frac{1-t}{2\pi}\sum_{n=0}^{\infty} t^n \sqrt{Q}\,\rho_{n;1}\,  \sqrt{Q}= \frac{1-t}{2\pi}\int_{\Pi_{+}}\mathrm{d}q\,\mathrm{d}p \,  \sqrt{Q}\,U(q,p)\, \sqrt{Q}\,\frac{1}{\sqrt{q}}\,\sum_{n=0}^{\infty}t^n\,U^{(\alpha)}_{nn}(q,p)\\
&=  \int_{\Pi_{+}}\mathrm{d}q\,\mathrm{d}p \,  \sqrt{Q}\,U(q,p) \,\sqrt{Q} \,\vap^{(\alpha)}_t(q,p)\,,
\end{align}
where the weight $\vap^{(\alpha)}_t(q,p)$ has an involved form issue from a generating function of Jacobi polynomials found in page 213 
of \cite{magnus66},
\begin{equation}
\label{Omegaqpt}
 \vap^{(\alpha)}_t(q,p)= \frac{1-t}{2\pi}\,\frac{2^{2\alpha + 1}\, q^{\frac{\alpha}{2}}}{\mathcal{R}(q,p;t)}\, \left( \frac{\bar Z_+}{Z_+}\right)^{\alpha + 1}
 \,\left(\bar Z_+ + t Z_+ + \mathcal{R}(q,p;t)\right)^{-\alpha}\, ,
\end{equation}
with
\begin{equation}
\label{calRqpt}
\mathcal{R}(q,p;t)= \left(\bar Z_+^2 -2 Y t \vert Z_+\vert^2 +t^2 Z_+^2 \right)^{1/2}\, .
\end{equation}
Finally note the integral representation of the simplest case $t=0$ which corresponds to the projector on the first basis element,
\begin{equation}
\label{rho0int}
\rho^{(\alpha)}_0= \mathsf{P}^{(\alpha)}_0 = |\eal_0\rg\lg\eal_0|=  \frac{2^{\alpha+1}}{2\pi}\int_{\Pi_{+}}\mathrm{d}q\,\mathrm{d}p \, \frac{q^{\alpha/2}}{(q+1 +2i q p)^{\alpha + 1}} \sqrt{Q}U(q,p) \sqrt{Q}\, . 
\end{equation}
This operator  is defined through its action in $L^2(\R_+^\ast, \ud x)$ by
\begin{equation}
\label{rho0s}
 (\rho^{(\alpha)}_0\psi)(x)= \frac{1}{\alpha! } x^{\frac{\alpha}{2}} e^{-\frac{x}{2}}\int_0^{+\infty}\ud y\, y^{\frac{\alpha}{2}}\,
 e^{-\frac{y}{2}}\, \psi(y)\,. 
\end{equation}

\section{Covariant affine integral quantization from weight}
\label{genform}
\subsection{ General results}
We now start from the framework of Section \ref{weight} and establish general formulas for quantization issued from 
a weight function $\vap(q,p)$, which obeys Eq.\;\eqref{eq:condition1} and so yields a symmetric operator  as
\begin{equation}
\int_{\Pi_{+}}\cdm^{-1} U(q,p)\cdm^{-1} \,\varpi(q,p)\,\mathrm{d}q\,\mathrm{d}p:=\sfMv\, . \label{affbopg}
\end{equation}
Let us first establish the nature of $\sfMv$ as an integral operator in $\mathcal{H}=L^{2}(\mathbb{R}_{+}^{\ast},\mathrm{d}x)$. 
\beprop
 The action on $\phi$ in  $\mathcal{H}$ of the operator $\sfMv$ defined by the integral representation \eqref{affbopg}  is given by 
\begin{equation}
\label{acMom}
(\sfMv \phi)(x) = \int_{0}^{\infty}\mathcal{M}^{\vap}(x,x^{\prime})\,\phi(x^{\prime})\,\ud x^{\prime}\, , 
\end{equation}
where the kernel $\cMv$ is given by
\begin{equation}
\label{kerMom}
\mathcal{M}^{\vap}(x,x^{\prime}) = \frac{1}{\sqrt{2\pi}}\,\frac{x}{x^{\prime}}\,\hat{\vap}_p\left(\frac{x}{x^{\prime}}, -x\right)\, . 
\end{equation}
Here $\hat{\vap}_p$ is the partial Fourier transform of $\vap$ with respect to the variable $p$, as it was defined  by Eq.\;\eqref{parcoure}.
\enprop
\bprf Let $\phi_1$, $\phi_2$ be two elements of $\calH$. Supposing that the expression $\lg \phi_1 | \sfMv|\phi_2\rg$ is finite, we have from  the action on the right of $U(q,p)$ and of the Duflo-Moore operators
\begin{align*}
\lg \phi_1 | \sfMv|\phi_2\rg &= \frac{1}{2\pi}\int_0^{+\infty} \frac{\ud q}{q} \int_0^{+\infty} \ud x \, x \, \overline{\phi_1(x)}\, \phi_2\left(\frac{x}{q}\right)\int_{-\infty}^{+\infty} \ud p\, e^{\ii px}\, \vap(q,p)\\
&= \frac{1}{\sqrt{2\pi}}\int_0^{+\infty} \frac{\ud q}{q} \int_0^{+\infty} \ud x \, x \, \overline{\phi_1(x)}\, \phi_2\left(\frac{x}{q}\right) \, \hat{\vap}_p(q,-x)\\
&= \frac{1}{\sqrt{2\pi}}\int_0^{+\infty} \ud x \int_0^{+\infty}\ud x^{\prime} \overline{\phi_1(x)}\, \phi_2(x^{\prime}) \, \frac{x}{x^{\prime}}\, \hat{\vap}_p\left(\frac{x}{x^{\prime}},-x\right)\, , 
\end{align*} 
the last equation being obtained through the change of variables $q\mapsto x^{\prime} = x/q$. 
\eprf

 We note that, if we impose $\sfMv$ to be symmetric operator, the resulting symmetry of the kernel
\begin{equation}
\label{symMv}
\cMv(x,x^{\prime})= \overline{\cMv(x^{\prime},x)} 
\end{equation}
is  trivially derived from Eq.\;\eqref{eq:condition1} and basic properties of the Fourier transform. 
\begin{coro}
Let the operator $\sfMv$ be a pure state $|\psi\rg\lg \psi|$ as it is for the construction of affine coherent states. Then  the corresponding weight function is given through its partial Fourier transform by
\begin{equation}
\label{acsvap}
\hat{\vap}_p(u,v)= \sqrt{2\pi}\, \frac{1}{u}\, \psi(-v)\,\overline{\psi\left(-\frac{v}{u}\right)}\, , \quad u>0\, , v<0\,. 
\end{equation}
In particular, we have the relation for the modulus of $\psi$
\begin{equation}
\label{om1-x}
\hat{\vap}_p(1,-x)=\sqrt{2\pi} \,\vert \psi(x)\vert^2\,. 
\end{equation}
\end{coro}
\bprf 
Immediate from 
\begin{equation*}
\cMv(x,x^{\prime})= \psi(x)\,\overline{\psi(x^{\prime})}\,. 
\end{equation*}
\eprf 

Let us now carry out the integral quantization \eqref{quantgr} with $G= \mathrm{Aff}_+(\R)$ and $\sfM= \sfMv$:
\begin{equation}
\label{genqvap}
f\mapsto A^{\vap}_f= \int_{\Pi_+}\frac{\ud q\,\ud p}{c_{\sfMv}} \, f(q,p)\, \sfMv(q,p)\, .
\end{equation}
By construction, the quantization map \eqref{genqvap} is covariant with respect to the unitary
affine action $U$:
\begin{equation}
\label{covaff} U(q_0,p_0) A^\vap_f U^{\dag}(q_0,p_0) =
A\vap_{\mathfrak{U}(q_0,p_0)f}\, ,
\end{equation}
with
\begin{equation}
\label{covaff2}
 \left(\mathfrak{U}(q_0,p_0)f\right)(q,p)=
f\left((q_0,p_0)^{-1}(q,p)\right)= f\left(\frac{q}{q_0},q_0(p
-p_0) \right)\, ,
\end{equation}
$\mathfrak{U}$ being the left regular representation of the affine
group.
\beprop
The action on $\phi$ in $\mathcal{H}$ of the operator $A^{\vap}_f$ defined by the integral quantization map \eqref{genqvap}  is given by 
\begin{equation}
\label{acMom}
(A^{\vap}_f\phi)(x) = \int_{0}^{+\infty}\mathcal{A}^{\vap}_f(x,x^{\prime})\,\phi(x^{\prime})\,\ud x^{\prime}\, , 
\end{equation}
where the kernel $\mathcal{A}^{\vap}_f$ is defined as
\begin{align}
\label{kerMomA}
\mathcal{A}^{\vap}_f(x,x^{\prime}) &= \frac{\sqrt{2\pi}}{c_{\sfMv}}\int_0^{+\infty}\frac{\ud q}{q}\,\cMv\left(\frac{x}{q},\frac{x^{\prime}}{q}\right)\, \hat{f}_p(q,x^{\prime}-x)\\
\label{kerMomB}  &= \frac{1}{c_{\sfMv}}\, \frac{x}{x^{\prime}}\int_0^{+\infty}\frac{\ud q}{q}\,\hat\vap_p\left(\frac{x}{x^{\prime}},-q\right)\, \hat{f}_p\left(\frac{x}{q},x^{\prime}-x\right)\, .
\end{align}
Here $\hat{f}_p$ is the partial Fourier transform of $f$ with respect to the variable $p$ defined in Eq.\;\eqref{parcoure}.
\enprop
\bprf Let $\phi_1$, $\phi_2$ be two elements of $\calH$. Supposing that the expression $\lg \phi_1 | A^{\vap}_f\ |\phi_2\rg$ makes sense, we have from  the action  of $U^{\dag}(q,p)$ on the right and  of $U(q,p)$ on the left,
\begin{align*}
\lg \phi_1 | A^{\vap}_f|\phi_2\rg &= \frac{1}{c_{\sfMv}}\int_0^{+\infty} \ud q \,q\int_0^{+\infty} \ud x \int_0^{+\infty} \ud x^{\prime}  \, \overline{\phi_1(qx)}\,\cMv(x,x^{\prime})\, \phi_2(qx^{\prime})\times \\&\times \int_{-\infty}^{+\infty} \ud p\, e^{-\ii qp(x^{\prime}-x)}\, f(q,p)\\
&= \frac{\sqrt{2\pi}}{c_{\sfMv}} \int_0^{+\infty} \ud q \,q\int_0^{+\infty} \ud x \int_0^{+\infty} \ud x^{\prime}  \, \overline{\phi_1(qx)}\times\\ \nonumber &\quad \times \cMv(x,x^{\prime})\, \phi_2(qx^{\prime})\hat{f}_p(q,q(x^{\prime}-x))\\
&=  \frac{\sqrt{2\pi}}{c_{\sfMv}} \int_0^{+\infty} \ud x \int_0^{+\infty} \ud x^{\prime} \overline{\phi_1(x)}\,\phi_2(x^{\prime})\int_0^{+\infty} \frac{\ud q}{q}\,   \cMv\left(\frac{x}{q},\frac{x^{\prime}}{q}\right)\,\hat{f}_p(q,x^{\prime}-x)\, , 
\end{align*} 
the last equation being obtained through the changes of variables $qx\mapsto x$ and $qx^{\prime}\mapsto x^{\prime}$. 
Eq.\;\eqref{kerMomB} is obtained by using \eqref{kerMom} and the change $q\mapsto x/q$. 
\eprf
From the general form \eqref{kerMomB} of the integral kernel and particularizing to $f=1$, i.e. $\hat f_p\left(\frac{x}{q},x^{\prime}-x\right)= \sqrt{2\pi}\,\delta(x^{\prime}-x)$, case which corresponds to  the resolution of the identity, we get the relation between the normalisation constant $c_{\sfMv}$ and an integral on the weight function. 
\begin{coro}
A necessary condition for having the resolution of the identity issued from a choice of weight function $\vap(q,p)$ is 
\begin{equation}
\label{cnresun}
c_{\sfMv} =  \sqrt{2\pi} \int_0^{+\infty}\frac{\ud q}{q}\,\hat\vap_p\left(1,-q\right) \equiv  \sqrt{2\pi} \, \Omega(1)< \infty\,, 
\end{equation}
where we have introduced the function 
\begin{equation}
\label{Omu}
\Omega(u)= \int_0^{+\infty}\frac{\ud q}{q}\,\hat\vap_p\left(u,-q\right)\,. 
\end{equation}
\end{coro}
\subsection{Particular cases}
\subsubsection*{Position dependent functions $f$}
Suppose that $f$ does not depend on $p$, $f(q,p)\equiv u(q)$. From 
\begin{equation*}
\hat{f}_p(q,x^{\prime}-x)= \sqrt{2\pi}\,u(q)\,\delta(x^{\prime}-x)\, , 
\end{equation*}
and after integration one obtains for \eqref{kerMomA}
\begin{align}
\label{Auq}
\nonumber
\mathcal{A}^{\vap}_u(x,x^{\prime})&= \frac{2\pi}{c_{\sfMv}}\,\delta(x-x^{\prime})\int_0^{+\infty}\frac{\ud q}{q}\,\cMv\left(\frac{x}{q},\frac{x}{q}\right)\,u(q)\\
 &= \frac{2\pi}{c_{\sfMv}}\,\delta(x-x^{\prime})\int_0^{+\infty}\frac{\ud q}{q}\,\cMv\left(q,q\right)\,u\left(\frac{x}{q}\right)\,.
\end{align}
 Thus,  the quantum version of the function $u(q)$  is the multiplication operator
 \begin{equation}
\label{quq}
A^{\vap}_{u(q)}=  \frac{2\pi}{c_{\sfMv}}\int_0^{+\infty}\frac{\ud q}{q}\,\cMv\left(q,q\right)\,u\left(\frac{Q}{q}\right)
=  \frac{\sqrt{2\pi}}{c_{\sfMv}}\int_0^{+\infty}\frac{\ud q}{q}\,\hat\vap_p(1,-q)\,u\left(\frac{Q}{q}\right) 
\end{equation}
i.e. the multiplication by the convolution on the multiplicative group $\R_+^{\ast}$ of $u(x)$ with $ \frac{\sqrt{2\pi}}{c_{\sfMv}}\,\hat\vap_p(1,-x)$.

An interesting more particular case is when $u$ is a simple power of $q$, say $u(q)= q^{\beta}$. Then we have \begin{equation}
\label{Aqbeta}
A^{\vap}_{q^{\beta}}=  \frac{\sqrt{2\pi}}{c_{\sfMv}}\int_0^{+\infty}\frac{\ud q}{q^{1+\beta}}\,\hat\vap_p(1,-q)\,Q^{\beta}\equiv \frac{d_{\beta}}{d_0}\, Q^{\beta}\, , 
\end{equation}
where we have introduced the convenient notation 
\begin{equation}
\label{dbeta}
d_{\beta}= \int_0^{+\infty}\frac{\ud q}{q^{1+\beta}}\,\hat\vap_p(1,-q)\, , 
\end{equation}
together with necessary conditions of convergence. Note that with this notation, 
\begin{equation}
\label{cMomd0}
c_{\sfMv}= \sqrt{2\pi} d_0\,. 
\end{equation}
\subsubsection*{Momentum dependent functions $f$}
Now suppose that $f$ does not depend on $q$, $f(q,p)\equiv v(p)$. The formula \eqref{kerMomB} simplifies to 
\begin{equation}
\label{kerMomC}
\mathcal{A}^{\vap}_{v(p)}(x,x^{\prime}) =  \frac{1}{c_{\sfMv}}\,\hat v(x^{\prime}-x)\, \frac{x}{x^{\prime}}\int_0^{+\infty}\frac{\ud q}{q}\,\hat\vap_p\left(\frac{x}{x^{\prime}},-q\right)= \frac{1}{c_{\sfMv}}\, \hat v(x^{\prime}-x)\,\frac{x}{x^{\prime}}\,\Omega\left(\frac{x}{x^{\prime}}\right) \, . 
\end{equation}

As a simple but important example, let us examine the case $v(p)= p^n$, $n\in \N$. From distribution theory
\begin{equation}
\label{pnfour}
\hat v(x^{\prime}-x) = \sqrt{2\pi}\,\ii^n \,\delta^{(n)}(x^{\prime}-x)\, ,   
\end{equation}
we derive the differential action of the operator $A^{\vap}_{p^n}$ in $\mathcal{H}$ as the polynomial in $P= -\ii d/dx$
\begin{equation}
\label{Apn}
A^{\vap}_{p^n}= \frac{\sqrt{2\pi}}{c_{\sfMv}}\, \sum_{k=0}^{n}\binom{n}{k}\, \left(-\ii\frac{d}{dx^{\prime}}\right)^{n-k}\,\frac{x}{x^{\prime}}\,\Omega\left.\left(\frac{x}{x^{\prime}}\right)\right\vert_{x^{\prime}=x}\, P^k = P^n + \cdots\, .
\end{equation}
In particular
\begin{equation}
\label{Ap1}
A^{\vap}_{p}= P + \frac{\ii}{x}\,\left\lbrack 1 + \frac{\Omega^{\prime}(1)}{\Omega(1)}\right\rbrack\, . 
\end{equation}
This operator is symmetric but has no self-adjoint extension \cite{reedsimon2}. 
Hence, from \eqref{Aqbeta} with $\beta = 1$,  the canonical commutation rule holds up to a factor which can be easily put equal to one through a rescaling of the weight function
\begin{equation}
\label{ccrom}
[A_q,A_p]=  \frac{d_{\beta}}{d_0}\, \ii \, I\, . 
\end{equation} 
For the kinetic energy we have 
\begin{equation}
\label{Ap1}
A^{\vap}_{p^2}= P^2 + \frac{2\ii}{Q}\,\left\lbrack 1 + \frac{\Omega^{\prime}(1)}{\Omega(1)}\right\rbrack \, P -  \frac{1}{Q^2}\,\left\lbrack 2 + 4 \, \frac{\Omega^{\prime}(1)}{\Omega(1)} +\frac{\Omega^{\prime\prime}(1)}{\Omega(1)}\right\rbrack\, . 
\end{equation}
This  symmetric operator  is essentially self-adjoint or not, depending on the strength of the (attractive or repulsive) potentiel $1/x^2$ \cite{reedsimon2}. With the choice of a weight function such that $-2 - 4 \, \frac{\Omega^{\prime}(1)}{\Omega(1)} - \frac{\Omega^{\prime\prime}(1)}{\Omega(1)}\geq3/4$, it is essentially self-adjoint and so quantum dynamics of the free motion on the half line is unique.

\subsubsection*{Separable functions $f$}
Finally,  suppose that $f$ is separable, i.e.  $f(q,p)\equiv u(q)\,v(p)$. The formula \eqref{kerMomB} simplifies to 
\begin{equation}
\label{PosMomA}
\mathcal{A}^{\vap}_{u(q)v(p)}(x,x^{\prime}) =  \frac{1}{c_{\sfMv}}\,\hat v(x^{\prime}-x)\, \frac{x}{x^{\prime}}\int_0^{+\infty}\frac{\ud q}{q}\,\hat\vap_p\left(\frac{x}{x^{\prime}},-q\right)\,u\left(\frac{x}{q}\right)\,.\end{equation}
The elementary example is the quantization of the function $qp$ which produces the integral kernel and its corresponding operator
\begin{align}
\nonumber\label{dilaom}
\mathcal{A}^{\vap}_{qp}(x,x^{\prime}) &=  \frac{\sqrt{2\pi}}{c_{\sfMv}}\,\ii\,\delta^{\prime}(x^{\prime}-x)\, \frac{x^2}{x^{\prime}}\int_0^{+\infty}\frac{\ud q}{q^2}\,\hat\vap_p\left(\frac{x}{x^{\prime}},-q\right),\\
A^{\vap}_{qp}&= \frac{\Omega_1(1)}{\Omega(1)}\,D  + \ii\,\left\lbrack  \frac{3}{2}\frac{\Omega_1(1)}{\Omega(1)} + \frac{\Omega^{\prime}_1(1)}{\Omega(1)}\right\rbrack\, ,
\end{align}
where $D= \frac{1}{2}(QP+PQ)$ is the dilation generator.  As one of the two generators
(with $Q$)  of the UIR $U$ of the affine group, it is essentially
self-adjoint, with continuous spectrum $\lambda \in \R$ and corresponding 
eigendistributions $x^{\frac{1}{2} + \ii \lambda}$. We have introduced in \eqref{dilaom} one more notation with 
\begin{equation}
\label{Ombeta}
\Omega_{\beta}(u)= \int_0^{+\infty}\frac{\ud q}{q^{1+\beta}}\,\hat\vap_p\left(u,-q\right)\, , \quad \Omega_{0}(u) = \Omega(u)\, . 
\end{equation}

\subsection{Semi-classical portraits}
Given a weight function $\vap(q,p)$ yielding a symmetric unit trace  operator $\sfMv$, we  define the semi-classical or lower symbol of an operator $A$ in $\calH$ as the function
\begin{equation}
\label{semclA}
\check{A}(q,p):=  \mathrm{Tr}\left(A\,U(q,p)\,\sfMv\,U^{\dag}(q,p)\right) = \mathrm{Tr}\left(A\,\sfMv(q,p)\right)\, . 
\end{equation}
When the operator $A$ is the affine integral quantized version of a classical $f(q,p)$ with the same weight $\vap$, we get the transform of the type \eqref{lowsymb}
 \begin{equation}
\label{lowsymbv}
f(q,p)\mapsto \check f(q,p) \equiv \check{A}^{\vap}(q,p)= \int_{\Pi_+}\frac{\ud q^{\prime} \, \ud p^{\prime}}{c_{\sfMv}}\, f\left(qq^{\prime},\frac{p^{\prime}}{q}+p\right)\, \mathrm{Tr}\left(\sfMv(q^{\prime},p^{\prime})\sfMv\right)\ . 
\end{equation}
Of course, this expression has the meaning of an averaging of the classical $f$ if the function 
\begin{align}
\nonumber\label{distvap}
 (q,p)\equiv g &\mapsto \frac{1}{c_{\sfMv}}\mathrm{Tr}\left(\sfMv(g)\sfMv\right)=\\ 
\nonumber =& \frac{1}{c_{\sfMv}}\int_{\Pi_+}\ud g_1\,\ \vap(g_1)\,\int_{\Pi_+}\ud g_2\,\vap\left(g_2\right)\times\\  &\times \mathrm{Tr}\left(U(g)\cdm^{-1} U(g_1)\cdm^{-1}U\left(g^{-1}\right)\cdm^{-1} U(g_2)\cdm^{-1}\right)
\end{align}
is a true probability distribution on the half-plane, i.e. is positive since we know from the resolution of the identity that its integral is 1. So a new trace formula, extending \eqref{tracecdmU}, is needed here. Explicitly,
\begin{equation}
\label{trmommom}
\mathrm{Tr}\left(\sfMv(q,p)\sfMv\right)= \frac{1}{2\pi q} \, \int_0^{+\infty} \ud x\int_0^{+\infty} \ud y \, e^{-\ii p(y-x)}\, \hat \vap_p\left(\frac{x}{y},-\frac{x}{q}\right)\,\,\hat \vap_p\left(\frac{y}{x},-y\right)\, . 
\end{equation}
Integrating this expression on $\Pi_+$ with the measure $\dfrac{\ud q\,\ud p}{c_{\sfMv}}$ and using \eqref{Tr2cond} (i.e., $d_1 = \mathrm{Tr}(\sfMv) = 1$) and \eqref{cnresun}, we get $1$, which means that $\check 1= 1$, as expected. 

\section{Quantization with affine CS and thermal states}
\label{QaffCSth}
\subsection{Quantization with ACS}
\label{QaffCS}
Let us first implement the integral quantization scheme described above by
restricting the method  to the specific case of rank-one density operator or
projector $\sfMv=|\psi\rangle\left\langle \psi\right|$ where $\psi$
is a unit-norm admissible state, i.e. is in 
$L^{2}(\mathbb{R}_{+}^{\dagger},\mathrm{d}x) \cap
L^{2}(\mathbb{R}_{+}^{\dagger},\mathrm{d}x/x)$ (such a $\psi$ is also called
``fiducial vector\textquotedblright{} or
``wavelet\textquotedblright{}). With the notations of the previous section, we have from \eqref{acsvap} and \eqref{om1-x}
\begin{equation}
\label{Omdbeta}
\Omega(u)= \frac{\sqrt{2\pi}}{u}  \int_0^{+\infty}\frac{\ud q}{q}\, \psi(q)\, \overline{\psi\left(\frac{q}{u}\right)}\, , \quad d_{\beta}=\sqrt{2\pi}\int_0^{+\infty}\frac{\ud q}{q^{1+\beta}}\,\vert \psi(q)\vert^2\, . 
\end{equation}
In particular, 
\begin{align}
\label{Ompsi1a}
\Om(1)&=d_0\, , \quad \Om^{\prime}(1)= - \Om(1)  -\sqrt{2\pi}\lg\psi^{\prime}|\psi\rg \,,\\
\label{Ompsi1b} \Om^{\prime\prime}(1)&= 2\Om(1) + 4\sqrt{2\pi}\lg\psi^{\prime}|\psi\rg + \sqrt{2\pi}\lg\psi^{\prime\prime}|Q|\psi\rg\, . 
\end{align}
Note that $\lg\psi^{\prime}|\psi\rg$ is purely imaginary and cancels for real $\psi$. 

Therefore, by applying the general formalism, we recover  a set of results already given in previous works, e.g. in  \cite{berdagama14}.  As was already pointed out after stating the orthogonality relations \eqref{orthaffine}, the action of the UIR operators $U(q,p)$ on $\psi$ produces all affine coherent states, i.e. wavelets, defined as $|q,p\rangle\
=U(q,p)|\psi\rangle$. Immediate examples of such vectors are $\psi(x) = \eal_m(x)$, i.e., with $\rho_m = \mathsf{P}_m$,  for $\alpha \geq \alpha_0 >0$, where $\alpha_0$ is suitably chosen in view of quantizing a certain class of function $f(q,p)$. 

Hence,  to the irreducibility of the representation $U$ and its square-integrability  expressed by \eqref{orthaffine},
the corresponding quantization reads as
\begin{equation}
\label{quantfaff} f\ \mapsto\
A_{f}=\int_{\Pi_{+}}f(q,p)|q,p\rangle\langle
q,p|\dfrac{\mathrm{d}q\mathrm{d}p}{2\pi c_{-1}}\,,
\end{equation}
which arises from the resolution of the identity
\begin{equation}
\label{affresunit} \int_{\Pi_{+}}|q,p\rangle\langle
q,p|\,\dfrac{\mathrm{d}q\mathrm{d}p}{2\pi c_{-1}}=I\,,
\end{equation}
where we adopt for convenience the notations of \cite{berdagama14},
\begin{equation}
\label{cgamma}
c_{\gamma}:=\int_{0}^{\infty}|\psi(x)|^{2}\,\frac{\mathrm{d}x}{x^{2+\gamma}}= \frac{1}{\sqrt{2\pi}}\, d_{\gamma +1}\,.
\end{equation}
Thus, a necessary condition  to have \eqref{affresunit} true is
that $c_{-1} < \infty$, which implies $\psi(0) = 0$, a well-known
requirement in wavelet analysis.

To simplify, we choose a real fiducial vector. Then, 
\begin{equation}
\label{quantqp} A_{p}=P\,,\quad
A_{q^{\beta}}=\frac{c_{\beta-1}}{c_{-1}} \,Q^{\beta}\,.
\end{equation}
Whereas $Q$ is self-adjoint, we recall that the operator $P$ is symmetric but has no
self-adjoint extension.  The quantization of the product $qp$ yields:
\begin{equation}
\label{quantqpdil} A_{qp}= \frac{c_0}{c_{-1}}\frac{Q P + PQ}{2}
= \frac{c_0}{c_{-1}}\, D\,.
\end{equation}

The quantization of kinetic energy gives
\begin{equation}
\label{qkinener} A_{p^{2}}=P^{2}+KQ^{-2}\,,\quad
K=K(\psi)=\int_{0}^{\infty}
(\psi^{\prime}(u))^{2}\,u\frac{\mathrm{d}u}{c_{-1}}.
\end{equation}
Therefore, wavelet quantization prevents a quantum free particle
moving on the positive line from reaching the origin. As already discussed in the previous section,   the above regularized operator, defined on the domain of
smooth function of compact support, is essentially self-adjoint
for $K\geq3/4$ \cite{GezKiR85}, and then quantum dynamics of the free motion on the half line is unique. Whilst canonical quantization, based on Weyl-Heisenberg symmetry which is unnatural in the present case,   introduces ambiguity on the quantum level, ACS quantization with suitable fiducial vector removes this ambiguity.  

 The quantum states and their dynamics have semi-classical phase space
representations through symbols. For the state
$|\phi\rangle$ the corresponding symbol reads
\begin{equation}
\label{Phisym} \Phi(q,p)=\langle q,p|\phi\rangle/\sqrt{2\pi}\,,
\end{equation}
with the associated probability distribution on phase space given
by
\begin{equation}
\label{rhophi} \rho_{\phi}(q,p)=\dfrac{1}{2\pi c_{-1}}|\langle
q,p|\phi\rangle|^{2}.
\end{equation}
Having the (energy) eigenstates of some quantum Hamiltonian $\mathsf{H}$ at
our disposal, the most natural being in this context the quantized $A_h$ of a classical Hamiltonian $h(q,p)$, we can compute the time evolution
\begin{equation}
\label{rhophiev} \rho_{\phi}(q,p,t):=\dfrac{1}{2\pi
c_{-1}}|\langle q,p|e^{-\ii \mathsf{H}t}|\phi\rangle|^{2}
\end{equation}
for any state $\phi$. 

The map \eqref{lowsymb} yielding lower
symbols from classical $f$ reads in the present case (supposing
that Fubini holds):
\begin{align}
\label{afflowsymb} \nonumber \check{f}(q,p)=
\frac{1}{\sqrt{2\pi}c_{-1}}&\int_0^{\infty} \frac{\ud
q^{\prime}}{qq^{\prime}}\,\int_0^{\infty}\ud x
\,\int_0^{\infty}\ud
x^{\prime} e^{\ii p(x^{\prime}-x)}\times \\
& \times
\hat{f}_p(q^{\prime},x^{\prime}-x)\,\psi\left(\frac{x}{q}\right)\,
\psi\left(\frac{x}{q^{\prime}}\right)\,\psi\left(\frac{x^{\prime}}{q}\right)\,
\psi\left(\frac{x^{\prime}}{q^{\prime}}\right)\,,
\end{align}
where $\hat{f}_p$ stands for the partial inverse Fourier transform introduced in \eqref{parcoure}.

For functions $f$ depending on $q$ only, expression
\eqref{afflowsymb} simplifies to a lower symbol  depending on $q$
only:
\begin{equation}
\label{lowfq} \check{f}(q)=
\frac{1}{c_{-1}}\int_0^{\infty}\frac{\ud
q^{\prime}}{qq^{\prime}}\, f(q^{\prime}) \int_0^{\infty}\ud
x\,\psi^2\left(\frac{x}{q}\right)\,\psi^2\left(\frac{x}{q^{\prime}}\right)\,.
\end{equation}
For instance, any power of $q$ is transformed into the same power
up to a constant factor
\begin{equation}
\label{powq} q^{\beta} \mapsto  \check{q^{\beta}}=
\frac{c_{-\beta-1}c_{-\beta-2}}{c_{-1}} \, q^{\beta}\, .
\end{equation}
Note that $c_{-2} = 1$ from the normalisation of $\psi$. Other
important symbols are:
\begin{equation}
\label{symbp} p \mapsto \check{p}= p\, ,
\end{equation}
\begin{equation}
\label{symbp2} p^2 \mapsto \check{p^2}= p^2 +
\frac{\mathrm{c(\psi)}}{q^2}\, , \quad \mathrm{c(\psi)} =
\int_0^{\infty}\left(\psi^{\prime}(x)\right)^2\,(1+c_1x)\,\ud x\,.
\end{equation}
\begin{equation}
\label{symbp2A} qp \mapsto \check{qp}= \frac{c_0
c_{-3}}{c_{-1}}qp\,
\end{equation}
Another interesting formula in the  semi-classical context
concerns the Fubini-Study metric derived from the symbol of total
differential $d$ with respect to parameters $q$ and $p$ affine
coherent states,
\begin{equation}
\label{dercs} \lg q,p|d|q,p\rg= \ii q\,
dp\int_0^{\infty}(\psi(x))^2\, x\, \ud x = \ii q\, dp \,c_{-3} \,.
\end{equation}
and from norm squared of $d|q,p\rg$,
\begin{equation}
\label{norm2} \Vert d|q,p\rg\Vert^2= c_{-4}\,q^2\, dp^2 + L\,
\frac{dq^2}{q^2}\, , \quad L = \int_0^{\infty}\ud x\, x^2\,
(\psi^{\prime}(x)^{\prime})^2 - \frac{1}{4}\,.
\end{equation}
With Klauder's notations \cite{klauderscm}
\begin{equation}
\label{fubstud} d\sigma^2(q,p):= 2\left\lbrack \Vert d
|q,p\rg\Vert^2 - \vert \lg q,p|d|q,p\rg\vert^2\right\rbrack =
2\left((c_{-4}-c_{-3}^2)  q^2\, dp^2 + L\,
\frac{dq^2}{q^2}\right)\,.
\end{equation}

\subsection{Quantization of basic observables with Laguerre thermal state}
\label{thermal}

In this subsection we compute the quantized version of $f(q,p)$, e.g. the momentum $p$, the classical dilation $qp$, the kinetic energy $p^2$,  the power potential $q^{\beta}$ when the affine integral quantization is carried out with the thermal density operator $\rho_t\left(q,p\right)$.  Because the closed formula \eqref{Omegaqpt} for the weight function $\vap_\mathrm{th}(q,p;t)$ is quite intricate, it is more tractable to work directly with the expansion \eqref{plboltrhohp} of $\rho_t$, to use the ACS quantization formulae above for each rank one operator in the series and to sum the results. With the notations \eqref{residrhoTqpF}, \eqref{diagosol}, and \eqref{thermint},  the general formula reads as
\begin{equation}
\label{genthermal}
A^{\vap^{(\alpha)}_t}_f
= \frac{\alpha}{2\pi}\int_{\Pi_{+}}\rho^{(\alpha)}_t(q,p) \,f(q,p)\,\mathrm{d}q\,\mathrm{d}p= \frac{\alpha\,(1-t)}{2\pi}\sum_{n=0}^{\infty} t^n \, c^{(\alpha)}_{-1;n} \, A^{\vap^{(\alpha)}_n}_f\, , 
\end{equation}
where the constants
\begin{equation}
\label{cgamn}
 c^{(\alpha)}_{\gamma;n} := \frac{n!}{(n+\alpha)!}\int_0^{+\infty}\frac{\ud x}{x^{2+\gamma}}\,e^{-x}\,x^{\alpha}\left(\Lal_n(x)\right)^2
\end{equation}
can easily be deduced from 7.414 Eq. (12) in \cite{gradryz07}. 
\beprop
The  affine integral quantizations with Laguerre thermal state of the classical momentum, the kinetic energy, the dilation function $qp$, and  the powers of $q$  are  given by
\begin{equation}
\label{Apthe}
A^{\vap^{(\alpha)}_t}_p = P\, . 
\end{equation}
\begin{equation}
\label{Aqbetathe}
A^{\vap^{(\alpha)}_t}_{q^{\beta}} =c^{(\alpha)}_{\beta-1}(t)\, Q^{\beta}\, , \quad  c^{(\alpha)}_{\gamma}(t):=  \frac{\alpha\,(1-t)}{2\pi}\sum_{n=0}^{\infty}t^n\, c^{(\alpha)}_{\beta-1;n}\,.  
\end{equation}
\begin{equation}
\label{Aqpthe}
A^{\vap^{(\alpha)}_t}_{qp} =c^{(\alpha)}_{0}(t)\,D\, . 
\end{equation}
\begin{equation}
\label{Ap2the}
A^{\vap^{(\alpha)}_t}_{p^2} =P^2 + \frac{K^{(\alpha)}(t)}{Q^2}\, , \quad  K^{(\alpha)}(t):= \frac{\alpha\,(1-t)}{2\pi}\sum_{n=0}^{\infty}t^n\, \int_{0}^{\infty}
\left(\frac{d\eal_n(x)}{dx}\right)^{2}\,x\,\mathrm{d}x\,. 
\end{equation}
\enprop 
\bprf
These formulas are proved by the fact that all fiducial Laguerre basis elements are real and direct application of Eqs.\;\eqref{quantqp}\;\eqref{qkinener} and \eqref{qkinener}. 
\eprf

\section{The Inverse Affine Operator and Quantization of Basic Operators} 
\label{inversion}

In this section we investigate the integral quantization yielded by $\sfMa$. This operator is equal to twice the inversion operator $\mathcal{I}$ defined on $L^2(\R_+^*,\ud x)$ as it was  introduced in Eq.\;\eqref{invop}.

\subsection{Properties of the inversion map and the related quantization}

First of all let us return to the determination of the trace of the inversion operator $\mathcal{I}$. 
\beprop
The operator $\sfMa= 2\, \mathcal{I}$ is unit trace. 
\enprop
\bprf Although we have already proved this property by  application of Proposition \ref{neccondtr1}, it is useful to present here a direct proof through the use of the  orthonormal  Laguerre basis with $\alpha=0$.  One gets successively
\begin{align*}
 \mathrm{Tr}\left(\mathcal{I}\right)
 &=\sum_{n=0}^{\infty}\lg e^{(0)}_n|\mathcal{I}e^{(0)}_n\rg
= \sum_{n=0}^{\infty}\int_{0}^{\infty} \frac{\ud x}{x}e^{-\frac{x}{2}}e^{-\frac{1}{2x}}L_n\left(x\right)L_n\left(\frac{1}{x}\right)\\&= \int_{0}^{\infty} \frac{\ud x}{x}\sum_{n=0}^{\infty} e^{-\frac{x}{2}}e^{-\frac{1}{2x}}L_n\left(x\right)L_n\left(\frac{1}{x}\right)=  \int_{0}^{\infty} \frac{\ud x}{x}\delta\left(x-\frac{1}{x}\right)\\ &
 = \int_{0}^{\infty} \frac{\ud x}{x}\frac{\delta\left(x-1\right)}{2}= \frac{1}{2}
 \end{align*}
\eprf

\beprop
\label{propqinv}
The integral kernel \eqref{kerMomB} of the quantization of a function $f(q,p)$ through the weight function $\vapa$ given in \eqref{vapweyl} has the following expression,
\begin{equation}
\label{intkerqI}
\mathcal{A}_f^{a\mathcal{W}}(x,x^{\prime}) = \frac{1}{\sqrt{2\pi}}\, \hat f_p\left(\sqrt{\frac{x^{\prime}}{x}}, x^{\prime}-x\right)\,. 
\end{equation}
\enprop
\bprf
The computation of the partial Fourier transform of the weight function $\vapa(q,p)= e^{-\ii \sqrt{q}p}/\sqrt{q}$ is immediate and yields
\begin{equation}
\label{pfourvapa}
\widehat{(\vapa)}_p(q,k)= \sqrt{2\pi}\, \frac{\delta (k+ \sqrt{q})}{\sqrt{q}}\, . 
\end{equation}
It follows for the integral $\Om(u)$  defined by \eqref{Omu}  and   the constant \eqref{cnresun}  the simple values
\begin{equation}
\label{ cMvapa}
\Om(u)= \frac{\sqrt{2\pi}}{u}\, , \quad c_{\vapa}= 2\pi\,. 
\end{equation} 
Then, Eq.\;\eqref{intkerqI} results from the integration with delta distribution. 
\eprf
We derive from Proposition \ref{propqinv} the following interesting results holding for this particular type of integral quantization.
\beprop
\label{propqinvpcuv}
\begin{itemize}
  \item[(i)] The quantization of a function of $q$, $f(q,p)= u(q)$ provided by the weight $\vapa$ is   $u(Q)$.
 \item[(ii)] Similarly, the quantization of a function of $p$, $f(q,p)= v(p)$ provided by the weight $\vapa$ is   $v(P)$ (in the general sense of pseudo-differential operators produced by Fourier transform).
  \item[(iii)] More generally, the quantization of a separable function $f(q,p)= u(q)\,v(p)$ provided by the weight $\vapa$ is the integral operator
  \begin{equation}
\label{Auvinvq}
\left(A^{a\mathcal{W}}_{u(q)v(p)}\,\psi\right)(x)= \frac{1}{\sqrt{2\pi}}  \int_0^{+\infty}\ud x^{\prime}\,\hat v(x^{\prime}-x)\, u\left(\sqrt{x\,x^{\prime}}\right)\, \psi(x^{\prime})\, . 
\end{equation}
 \item[(iv)] In particular, the quantization of $u(q)\,p^n$, $n\in \N$, yields the symmetric operator,
 \begin{equation}
\label{Apnuinvq}
A^{a\mathcal{W}}_{u(q)p^n}= \sum_{k=0}^n \binom{n}{k} \,(-\ii)^{n-k}u^{(n-k)}(Q)\, P^k\, , 
\end{equation}
and for the dilation,
\begin{equation}
\label{Apqinvq}
A^{a\mathcal{W}}_{qp}= D\, .
\end{equation}
\end{itemize}
\enprop
\bprf The proof is made through a direct application of Eq.\;\eqref{intkerqI} and application of elementary distribution theory.
 \eprf
Therefore, this affine integral  quantization is the exact counterpart of the Weyl-Wigner  integral quantization \cite{bergaz14} and can be termed as canonical as well.  However, the choice of such a procedure would lead to three difficulties, at least,
\begin{enumerate}
  \item Since the quantization of the kinetic energy of the free particle on the half-line is just 
  \begin{equation}
\label{aWp2}
A^{a\mathcal{W}}_{p^2}= P^2\, , 
\end{equation}
and thus does not produce an essentially self-adjoint operator, the corresponding  quantum dynamics depends on boundary conditions at the origin $x=0$. There exists an irreducible ambiguity since different physics are possible on the quantum level. 
  \item No classical singularity is cured on the quantum level since
  \begin{equation}
\label{aWuqvp}
   A^{a\mathcal{W}}_{u(q)}  = u(Q)\, , \qquad A^{a\mathcal{W}}_{v(p)}  = v(P)\, ,
\end{equation}
a feature of the Weyl-Wigner  integral quantization as well.
\item The semi-classical portraits of quantum operators along Eqs.\;\eqref{semclA} and \eqref{lowsymbv} cannot be given a probabilistic interpretation, a feature of the Weyl-Wigner  integral quantization as well. 
\end{enumerate}
The last point is developed below for rank-one operators  $|\psi\rg\lg \psi|$, i.e. pure states. 

\subsection{Affine Wigner-like quasi-probability}
The  affine Wigner-like quasi-probability $\mathcal{AW}_{\psi}$ corresponding to the  state $\phi$ is the application of the general expression \eqref{semclA} to the projector $|\phi\rg\lg \phi|$:
\begin{equation}
\label{AWpsi}
\mathcal{AW}_{\phi}(q,p)
:=\lg \phi| \sfMa(q,p)|\phi \rg= 2\int_{0}^{\infty}\ud x \overline{\phi\left(x\right)} e^{\ii p\left(x-\frac{q^2}{x}\right)}\frac{q}{x}\phi\left(\frac{q^2}{x}\right)\,.
\end{equation}
\beprop
Let us consider a pure state $\phi$ and the corresponding quasi-probability distribution $\mathcal{AW}_{\phi}(q,p)$. The latter verifies the following properties.
\begin{itemize}
\item[(i)] It is real
\begin{equation} 
\overline{\mathcal{AW}_{\phi}(q,p)}=\mathcal{AW}_{\phi}(q,p)\,.
\end{equation} 
\item[(ii)] It is a quasi-probability,
\begin{equation}
\int_{\Pi_+}\frac{\ud q\;\ud p}{2\pi}\mathcal{AW}_{\phi}(q,p)=1\,.
\end{equation}
\item[(iii)] It satisfies the correct marginalization with respect to  variables $q$ and  $p$ respectively,
\begin{align}
\int_{0}^{\infty}\frac{\ud q}{2\pi} \mathcal{AW}_{\phi}(q,p)
=|\hat\phi(p)|^2\,.
\end{align}
\begin{align}
\int_{-\infty}^{+\infty}\frac{\ud p}{2\pi} \mathcal{AW}_{\phi}(q,p) 
=|\phi(q)|^2\,.
\end{align}
\end{itemize}
\enprop
\bprf
(i) and (ii) are  direct consequences of the fact that operator $ \sfMa$ is symmetric and unit trace (due to the resolution of the identity). (iii) results from elementary integral calculus on \eqref{AWpsi} through change of variable $q\mapsto q^2/x$ and two Fourier transforms (by considering  that  the support of $\phi(x)$ is  included in the positive half-line). 
\eprf
A last result concerns the semi-classical portrait of the operator $A^{a\mathcal{W}}_{f}$ obtained from the expressions\eqref{lowsymbv} and \eqref{trmommom}.
\beprop
The map $f(q,p) \mapsto \check f(q,p) = \mathrm{Tr}\left(A^{a\mathcal{W}}_{f}\, \sfM^{a\mathcal{W}}(q,p)\right)$ yields the following lower symbol
\begin{equation}
\label{affWiglow}
\check f(q,p)  = \frac{1}{\sqrt{2\pi}}\int_{-\infty}^{+\infty}\ud u\, e^{\ii pu}  \frac{\hat f_p(q, u)}{\sqrt{1+ \frac{u^2}{4q^2}}}= \frac{2q}{\pi}\, K_0(2qp)\ast_pf(q,p)\, , 
\end{equation}
where $K_0$ is a modified Bessel function of the second kind, and $\ast_p$ is the convolution product with respect to the variable $p$. 

In particular, the lower symbol of the quantization of a function of $q$ alone, $f(q,p)= u(q)$,  is   $u(q)$. 
\enprop

Thus, there is no ``round trip'' $f(q,p) \mapsto A^{a\mathcal{W}}_{f} \mapsto \check f(q,p)=f(q,p)$ here, contrary to the Wigner map based on the Weyl-Heisenberg symmetry.   It is interesting to notice that whereas the lower symbol of $A^{a\mathcal{W}}_{p}= P$ is exactly $\check p = p$, the lower symbol of the $A^{a\mathcal{W}}_{p^2}= P^2$ is $\check{p^2} = p^2 + 1/(4q^2)$. Therefore, even though this affine Wigner quantization does not regularizes the kinetic energy on the quantum level, it does on the level of the Wigner quasi-distribution. 
\subsection{Application to the half-oscillator}

\begin{figure}[htb]
\begin{center}
\includegraphics[scale=.7]{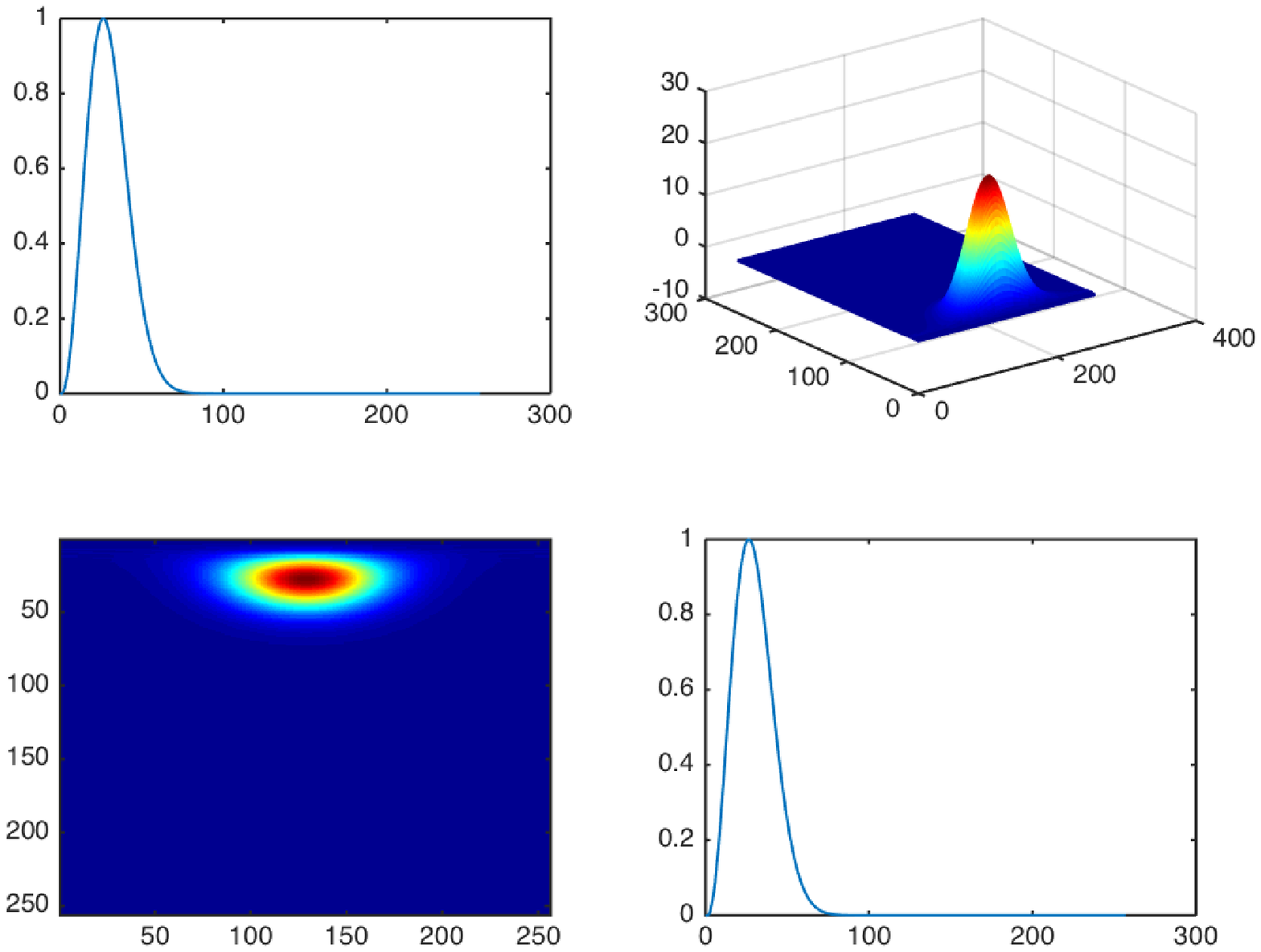}
\caption{First eigenfunction $\phi_{1}(x)=2x\;e^{-\frac{x^2}{2}}$: (a)-Plot of the probability density $|\phi_{1}(x)|^2$; (b)-3D plot of the corresponding Wigner quasi-density $\mathcal{AW}_{\phi_1}(q,p)$ ; (c)- 2D plot of $\mathcal{AW}_{\phi_1}(q,p)$; (d)-Sum over p which gives the reconstructed density $|\phi_{1}(q)|^2$}
\label{fig_wig1}
\end{center}
\end{figure}

\begin{figure}[htb]
\begin{center}
\includegraphics[scale=.7]{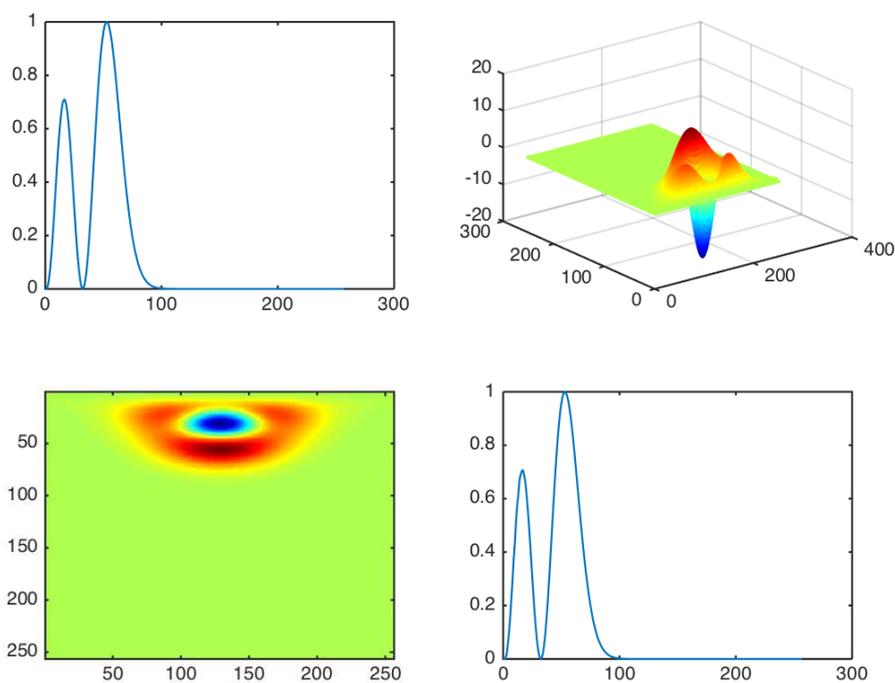}
\caption{Second eigenfunction $\phi_{2}(x)=2(3x^2-3x)\;e^{-\frac{x^2}{2}}$: (a)-Plot of the probability density $|\phi_{2}(x)|^2$; (b)-3D plot of the $\mathcal{AW}_{\phi_2}(q,p)$ ; (c)- 3D plot of  $\mathcal{AW}_{\phi_2}(q,p)$; (d)-Sum over p which gives the reconstructed density $|\phi_{2}(q)|^2$}
\label{fig_wig2}
\end{center}
\end{figure}

\begin{figure}[htb]
\begin{center}
\includegraphics[scale=.7]{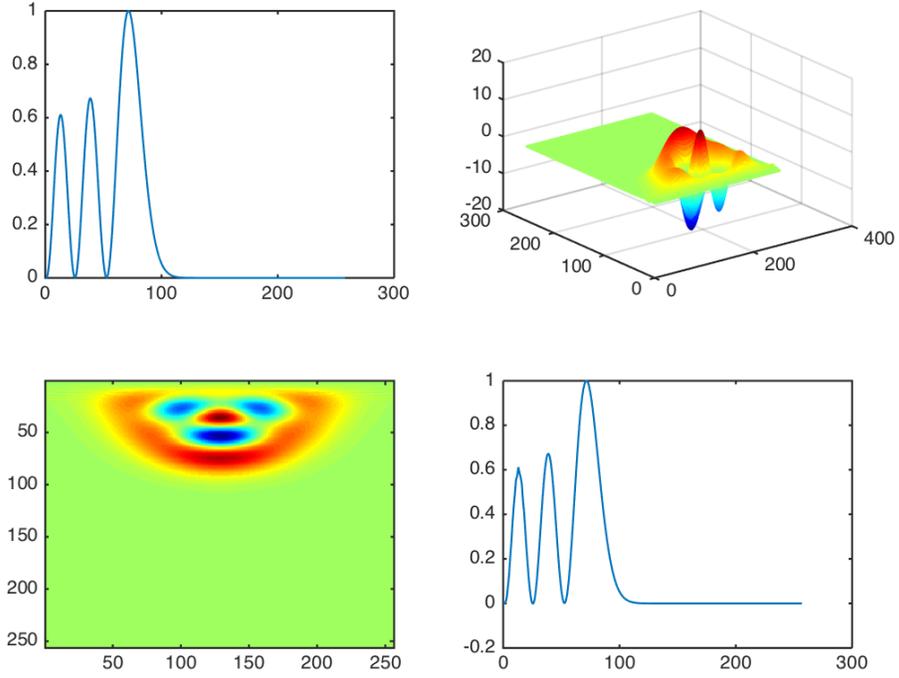}
\caption{Third eigenfunction $\phi_{3}(x)=8(4x^5-20x^3+15x)\;e^{-\frac{x^2}{2}}$: (a)-Plot of the probability density $|\psi_{3}(x)|^2$; (b)-3D plot of the $\mathcal{AW}_{\phi_3}(q,p)$ ; (c)- 3D plot of  $\mathcal{AW}_{\phi_3}(q,p)$; (d)-Sum over p which gives the reconstructed density $|\phi_{3}(q)|^2$}
\label{fig_wig3}
\end{center}
\end{figure}

\begin{figure}[htb]
\begin{center}
\includegraphics[scale=.7]{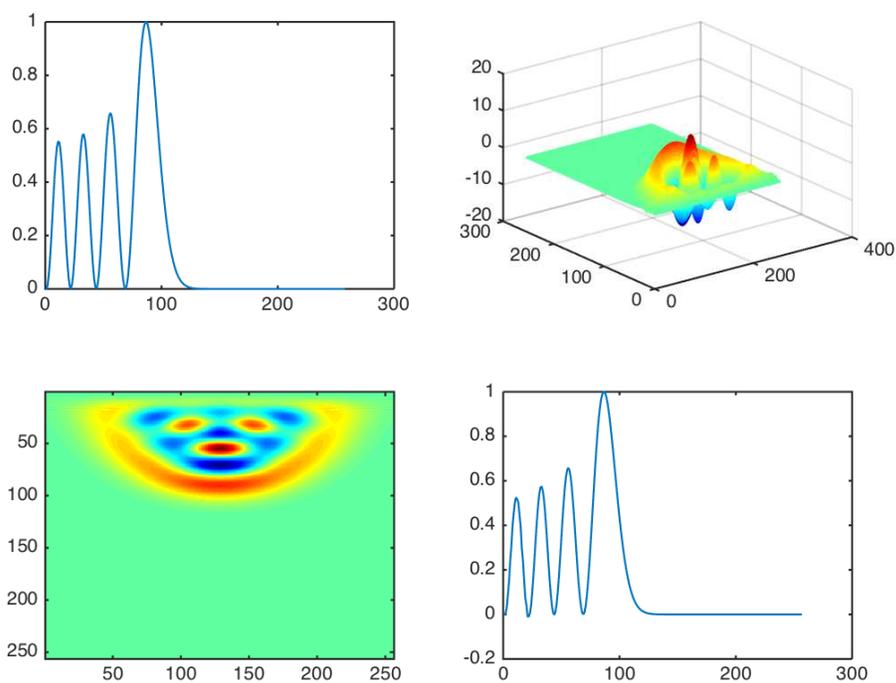}
\caption{Fourth eigenfunction $\phi_{4}(x)= (128 x^7-1344 x^5+3360 x^5+1334 x^3-1680 x)\;e^{-\frac{x^2}{2}}$: (a)-Plot of the probability density $|\phi_{4}(x)|^2$; (b)-3D plot of  $\mathcal{AW}_{\phi_4}(q,p)$ ; (c)- 3D plot of the $\mathcal{AW}_{\psi_4}(q,p)$; (d)-Sum over p which gives the reconstructed density $|\phi_{4}(q)|^2$}
\label{fig_wig4}
\end{center}
\end{figure}

\begin{figure}[htb]
\begin{center}
\includegraphics[scale=.7]{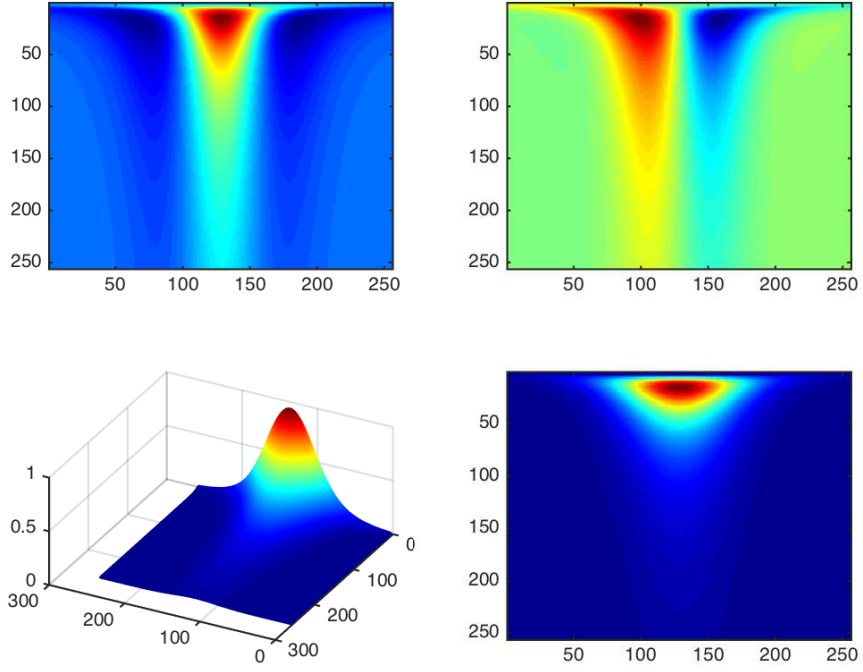}
\caption{First eigenfunction $\phi_{1}=2x\;e^{-\frac{x^2}{2}}$: (a)-2D plot of the real part of its wavelet transform $W_{\phi}(q,p)$; (b)- 2D plot of the imaginary part of $W_{\phi}(q,p)$; (c)- 3D plot of $\rho_{\phi}(q,p$ (d)-2D plot of $\rho_{\phi}(q,p)$.}
\label{fig_wave1}
\end{center}
\end{figure}

\begin{figure}[htb]
\begin{center}
\includegraphics[scale=.7]{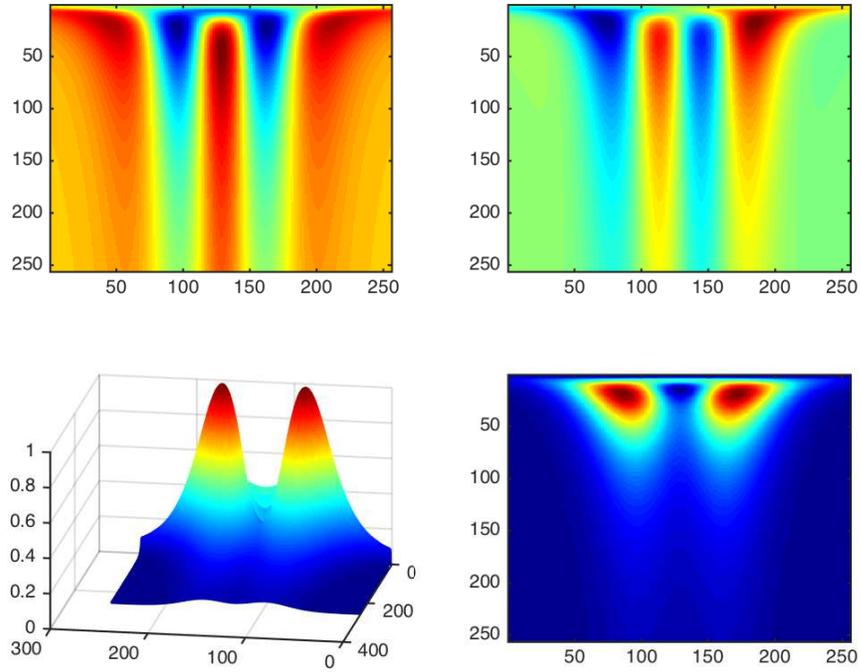}
\caption{Second eigenfunction $\phi_{2}=2(2x^3-3x)\;e^{-\frac{x^2}{2}}$: (a)-2D plot of the real part of $W_{\phi}(q,p)$; (b)- 2D plot of the imaginary part of $W_{\phi}(q,p)$; (c)- 3D plot of $\rho_{\phi}(q,p$ (d)-2D plot of $\rho_{\phi}(q,p)$.}
\label{fig_wave2}
\end{center}
\end{figure}

\begin{figure}[htb]
\begin{center}
\includegraphics[scale=.7]{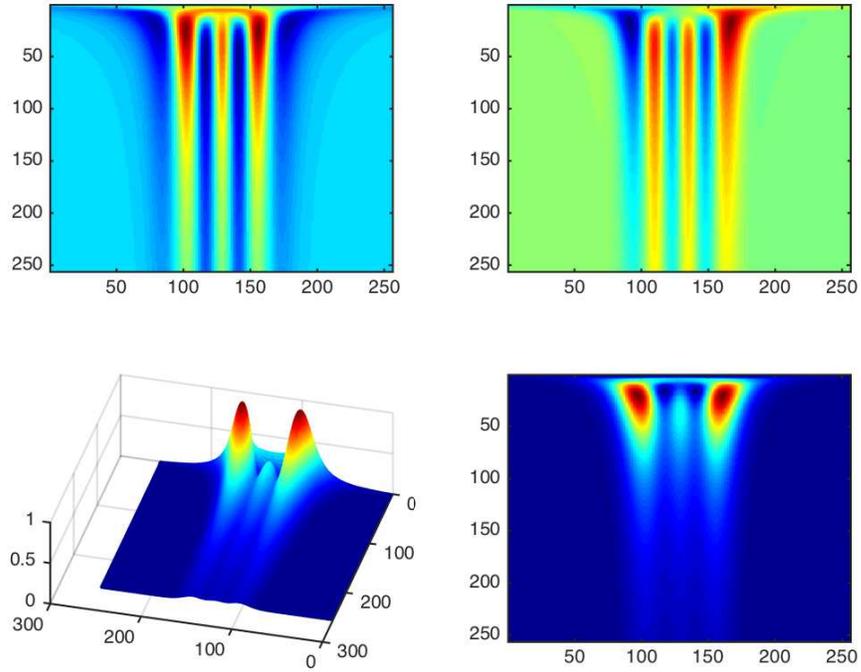}
\caption{Third eigenfunction $\phi_{3}=8(4x^5-20x^3+15x)\;e^{-\frac{x^2}{2}}$: (a)-2D plot of the real part of $W_{\phi}(q,p)$; (b)- 2D plot of the imaginary part of $W_{\phi}(q,p)$; (c)- 3D plot of $\rho_{\phi}(q,p$ (d)-2D plot of $\rho_{\phi}(q,p)$.}
\label{fig_wave3}
\end{center}
\end{figure}

\begin{figure}[htb]
\begin{center}
\includegraphics[scale=.7]{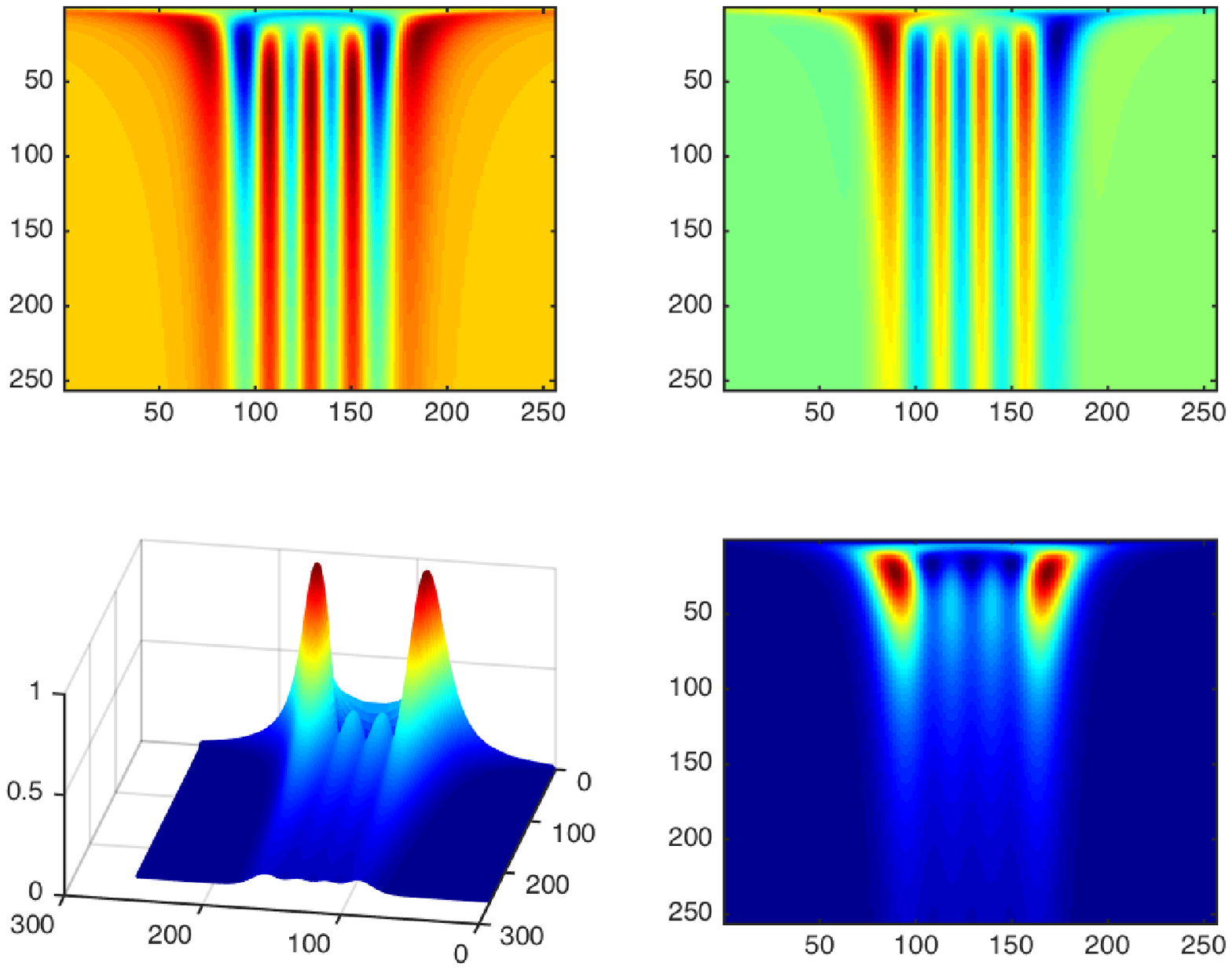}
\caption{Fourth eigenfunction $\phi_{4}= (128 x^7-1344 x^5+3360 x^5+1334 x^3-1680 x)\;e^{-\frac{x^2}{2}}$: (a)-2D plot of the real part of $W_{\phi}(q,p)$; (b)- 2D plot of the imaginary part of $W_{\phi}(q,p)$; (c)- 3D plot of $\rho_{\phi}(q,p)$ (d)-2D plot of $\rho_{\phi}(q,p)$.}
\label{fig_wave4}
\end{center}
\end{figure}

\begin{figure}[htb]
\begin{center}
\includegraphics[scale=.7]{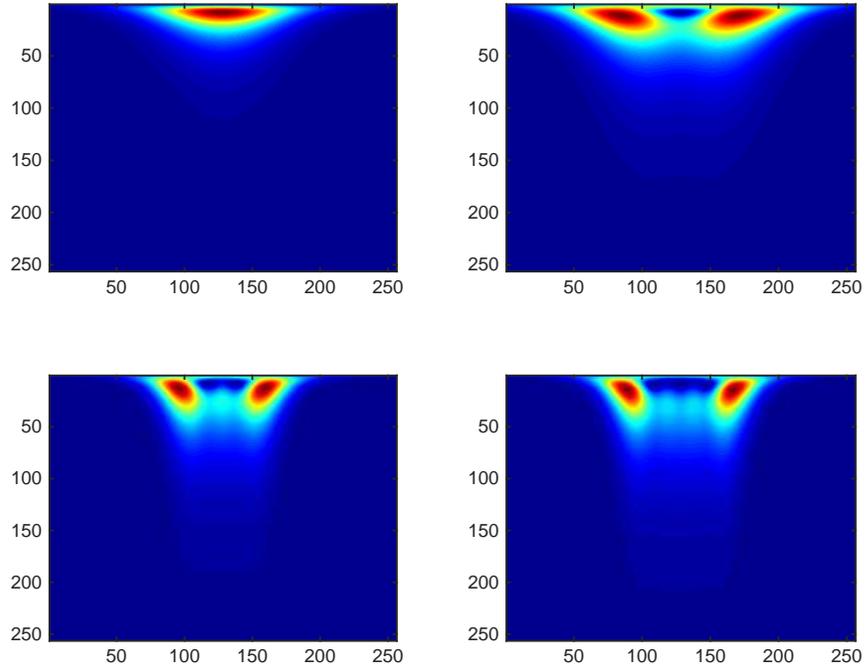}
\caption{2D plot of $\rho_{\phi}(q,p)$ for : (a) $\phi(x)=\phi_{1}(x)$; (b) $\phi(x)=\phi_{2}(x)$ ; (c) $\phi(x)=\phi_{3}(x)$; (d) $\phi(x)=\phi_{4}(x)$}
\label{fig_dens1}
\end{center}
\end{figure}

\begin{figure}[htb]
\begin{center}
\includegraphics[scale=.7]{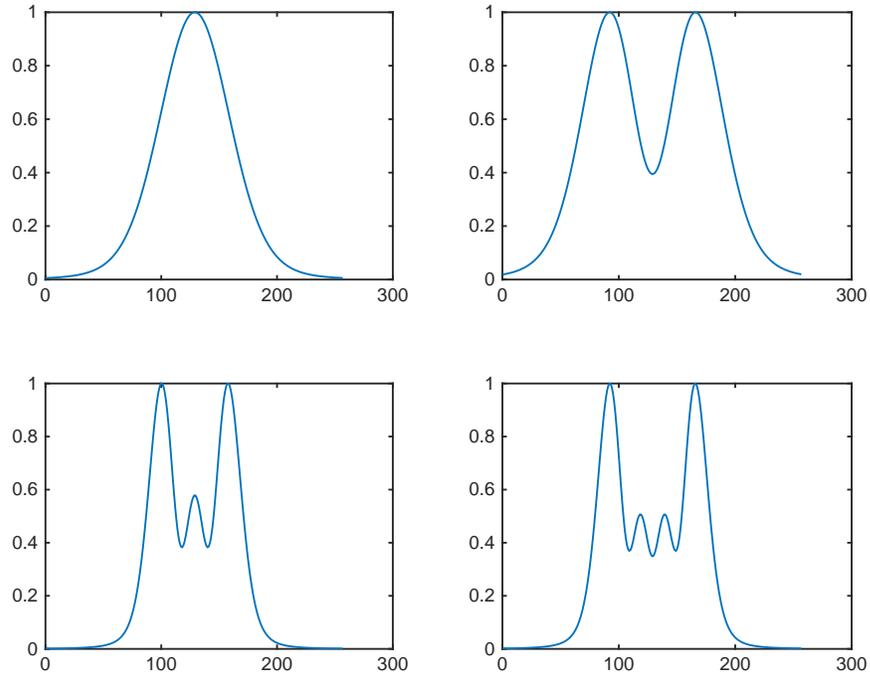}
\caption{1D plot of the sum over $q$ of $\rho_\phi(q,p)$, i.e, probability density in the momentum space for: (a) $\phi(x)=\phi_{1}(x)$; (b) $\phi(x)=\phi_{2}(x)$ ; (c) $\phi(x)=\phi_{3}(x)$; (d) $\phi(x)=\phi_{4}(x)$}
\label{fig_dens2}
\end{center}
\end{figure}

We consider the example of the half-harmonic  oscillator \cite{griffiths05}, that is, whose the motion is restricted to   the half-line. A physical interpretation of this could be a spring that can be stretched from its equilibrium position but not compressed. In this case where, for convenience, we put  $m=1$ (mass), $\omega=1$ (frequency), $\hbar=1$, the affine-Wigner quantization of the classical Hamiltonian $h_{1/2\mathrm{osc}}(q,p) = (p^2 + q^2)/2$ yields $A^{a\mathcal{W}}_{h} = \left(P^2 + Q^2\right)/2 \equiv H_{1/2\mathrm{osc}}$, which acts in $L^2({R_{+}}, \ud x)$ as the Shr\"{o}dinger operator 
\begin{equation}
\mathsf{H}_{1/2\mathrm{osc}}\,\phi(x):= \left(- \frac{1} {2}\,\frac{d^2} {dx^2}+ \frac{x^2} {2}\right)\phi(x)\, , \quad   x > 0\, .
\end{equation}
This operator is symmetric but not essentially self-adjoint. In solving the eigenvalue problem $\mathsf{H}_{\mathrm{sosc}}\,\phi = E\, \phi$,  it is necessary to choose Dirichlet boundary conditions such that the allowed solutions $\phi$  in $L^2({R_{+}})$  satisfy  $\phi(0)=0$. This yields  the odd Hermite functions as  allowed eigenfunctions, that is, the normalized
\begin{equation}
\label{psin}
\phi_{n}(x) = \pi^{-1/4}\frac{1}{2^n\sqrt{(2n-1)!}}H_{2n-1}(x) e^{-\frac{x^2} {2}}
\end{equation}
for $n=1,2,3,...$. 
As $\phi_n(x)$ is real the corresponding Wigner quasi-density is reduced to the simpler expression:
\begin{equation}
\label{AWpsi}
\mathcal{AW}_{\phi}(q,p)
= 2\int_{0}^{\infty}\ud x\, \phi\left(x\right)\, \cos{\left(p\left(x-\frac{q^2}{x}\right)\right)}\, \frac{q}{x} \,\phi\left(\frac{q^2}{x}\right)\,.
\end{equation}
In Figures \ref{fig_wig1}, \ref{fig_wig2}, \ref{fig_wig3}, \ref{fig_wig4} \ref{fig_wave1},\ref{fig_wave2},\ref{fig_wave3}, \ref{fig_wave4}, \ref{fig_dens1}, and  \ref{fig_dens2}, we give the 2D and 3D plots of the first four eigenfunctions together with their corresponding Wigner quasi-densities $\mathcal{AW}_{\phi}(q,p)$, their ACS symbols and probability densities  defined in terms of thermal states at $t=0$, i.e. ACS,  and for $\alpha = 1$, that is  
\begin{align}
\label{wavetr}
\nonumber \mathrm{Tr}\left(\rho_0^{(1)}(q,p)\,|\phi\rg\lg\phi|\right) &=\lg\phi|U(q,p)\rho^{(1)}_{0}U(q,p)^\dagger|\phi\rg\\
&= \left\vert\sqrt{\frac{\pi}{q}}\int^{+\infty}_{0}\;\ud x \;e^{\left(-\frac{1}{2q}+ip\right) x}\,\left( \frac{x}{q}\right)^{\frac{1}{2}}\, \phi(x)\right\vert^2\\\nonumber
&\equiv |W_{\phi}(q,p)|^2\equiv \rho_\phi(q,p)\, .
\end{align}
where $W_{\phi}(q,p)= \lg q,p|\phi\rg$ is the ACS symbol of $|\phi\rg$, equivalently the so-called \textit{wavelet transform} of  $\phi$ with respect to  $e^{(2)}_0$.

We notice the organization in negative and positive parts for $\mathcal{AW}_{\phi}(q,p)$ to be compared with the positive shape of $\rho_\phi(q,p)$.

\newpage

\section{Conclusion}
\label{conclu}
 In this paper we have explored the possibilities offered by affine covariant integral quantization beyond the familiar case of affine coherent states. The central object is the weight function $\vap(q,p)$ on the half-plane and its partial Fourier transform with respect to the momentum variable $p$. The half-plane itself was viewed here as the phase space for the motion of a point particle on the half-line. Actually it can be viewed as the phase space of a dynamical physical quantity which is positive, for which the value $0$ represents a singularity. This is the case for instance in cosmology with the volume of the Universe, the canonical conjugate being the expansion coordinate. It would be highly interesting to find other physical examples really accessible to observations, for instance in condensed matter physics, in order to favor this affine quantization with probabilistic content preferably  to the Weyl-Heisenberg canonical quantization  or the above affine Wigner one. In our sense, the essential self-adjointness of the quantum kinetic energy is a condition which should be always requested, together with a sound probabilistic interpretation of the semi-classical portraits of quantum operators issued from our quantization procedure. 
 
 An interesting problem to be addressed in this perpective is to find the class of weight functions $\varpi$ for which  the strength of the inverse square potential appearing in Eq.\;\eqref{Omu} is exactly $3/4$, i.e.
 \begin{equation}
\label{condkinsa}
\left\lbrack 2 + 4 \, \frac{\Omega^{\prime}(1)}{\Omega(1)} +\frac{\Omega^{\prime\prime}(1)}{\Omega(1)}\right\rbrack = -\frac{3}{4} \quad\mbox{with}\quad  \Omega(u)= \int_0^{+\infty}\frac{\ud q}{q}\,\hat\vap_p\left(u,-q\right)\,,
\end{equation} 
i.e. the lowest limit value for which unique self-adjointness of the quantum  kinetic term $P^2 + K/Q^2$ holds, and to compare that class of $\vap$'s with the class of $\vap$'s for which $\sfMv$ is a density operator giving rise to a real probabilistic interpretation. 

In our next work, we will develop a similar quantization approach based on the unitary irreducible representation of the  two-dimensional similitude group SIM$(2)$ \cite{alkramu00} for which  the four-dimensional phase space  is $\R^2 \times \R^2_\ast$, where $\R^2_\ast$ is the two-dimensional plane with the origin removed. Consistently to one of the main issues of the present work, our method yields a regularization of the singularity at the origin of the configuration plane on the quantum level and opens interesting opportunities in dealing with physical models presenting such point singularities.

\end{document}